\documentclass[prd,aps,showpacs,epsf,floats,onecolumn]{revtex4}%
\usepackage{amssymb}
\usepackage{amsfonts}
\usepackage{amsmath}
\usepackage{graphicx}
\usepackage{amssymb}
\usepackage{longtable}%
\providecommand{\U}[1]{\protect\rule{.1in}{.1in}}
\begin{document}
\title{\textbf{Transition Probabilities in Generalized Quantum Search Hamiltonian
Evolutions}}
\author{\textbf{Steven Gassner}$^{1}$\textbf{, Carlo Cafaro}$^{1}$\textbf{, and
Salvatore Capozziello}$^{2,3,4}$}
\affiliation{$^{1}$SUNY Polytechnic Institute, 12203 Albany, New York, USA}
\affiliation{$^{2}$Dipartimento di Fisica \textquotedblleft E. Pancini\textquotedblright,
Universit\`{a} di Napoli \textquotedblleft Federico II\textquotedblright,
I-80126, Napoli, Italy}
\affiliation{$^{3}$INFN Sez. di Napoli, Compl. Univ. di Monte S. Angelo, Edificio G,
I-80126, Napoli, Italy,}
\affiliation{$^{4}$Laboratory for Theoretical Cosmology, }
\affiliation{Tomsk State University of Control Systems and Radioelectronics (TUSUR), 634050
Tomsk, Russia.}

\begin{abstract}
A relevant problem in quantum computing concerns how fast a source state can
be driven into a target state according to Schr\"{o}dinger's quantum
mechanical evolution specified by a suitable driving Hamiltonian. In this
paper, we study in detail the computational aspects necessary to calculate the
transition probability from a source state to a target state in a continuous
time quantum search problem defined by a multi-parameter generalized
time-independent Hamiltonian. In particular, quantifying the performance of a
quantum search in terms of speed (minimum search time) and fidelity (maximum
success probability), we consider a variety of special cases that emerge from
the generalized Hamiltonian. In the context of optimal quantum search, we find
it is possible to outperform, in terms of minimum search time, the well-known
Farhi-Gutmann analog quantum search algorithm. In the context of nearly
optimal quantum search, instead, we show it is possible to identify
sub-optimal search algorithms capable of outperforming optimal search
algorithms if only a sufficiently high success probability is sought. Finally,
we briefly discuss the relevance of a tradeoff between speed and fidelity with
emphasis on issues of both theoretical and practical importance to quantum
information processing.

\end{abstract}

\maketitle

\section{Introduction}

Grover proposed a quantum algorithm for solving large database search problems
in Ref. \cite{grover97,grover01} . Grover's search algorithm helps searching
for an unknown marked item in an unstructured database of $N$ items by
accessing the database a minimum number of times. From a classical standpoint,
it is necessary to test $N/2$ items, on average, before finding the correct
item. With Grover's algorithm however, the same task can be completed
successfully with a complexity of order $\sqrt{N}$, that is, with a quadratic
speed up. Grover's algorithm was presented in terms of a discrete sequence of
unitary logic gates (digital quantum computation). Specifically, the
transition probability from the source state $\left\vert \psi_{s}\right\rangle
$ to the target state $\left\vert \psi_{w}\right\rangle $ after the $k$-times
sequential application of the so-called Grover quantum search iterate $G$ is
given by,%
\begin{equation}
\mathcal{P}_{\text{Grover}}\left(  k\text{, }N\right)  \overset{\text{def}}%
{=}\left\vert \left\langle \psi_{w}|G^{k}|\psi_{s}\right\rangle \right\vert
^{2}=\sin^{2}\left[  \left(  2k+1\right)  \tan^{-1}\left(  \frac{1}{\sqrt
{N-1}}\right)  \right]  \text{.} \label{pgrover}%
\end{equation}
In the limit of $N$ approaching infinity, $\mathcal{P}_{\text{Grover}}$ in Eq.
(\ref{pgrover}) approaches one if $k=O\left(  \sqrt{N}\right)  $. We point out
that the big $O$-notation $f\left(  x\right)  =O\left(  g\left(  x\right)
\right)  $ means that there exist\emph{ real} constants $c$ and $x_{0}$ such
that $\left\vert f\left(  x\right)  \right\vert \leq c\left\vert g\left(
x\right)  \right\vert $ for any $x\geq x_{0}$.

The temporal evolution of the state vector $\left\vert \psi\left(  t\right)
\right\rangle $ of a closed quantum system is characterized by the
Schr\"{o}dinger equation,%
\begin{equation}
i\hslash\partial_{t}\left\vert \psi\left(  t\right)  \right\rangle
=\mathcal{H}\left(  t\right)  \left\vert \psi\left(  t\right)  \right\rangle
\text{,} \label{H1}%
\end{equation}
where $\hslash\overset{\text{def}}{=}h/\left(  2\pi\right)  $ is the reduced
Planck constant, $i$ denotes the imaginary complex unit, and $\partial
_{t}\overset{\text{def}}{=}\partial/\partial t$. The Hamiltonian
$\mathcal{H}\left(  t\right)  $ in Eq. (\ref{H1}) encodes all relevant
information about the time evolution of the quantum system. From a quantum
computing standpoint, if the Hamiltonian $\mathcal{H}\left(  t\right)  $ is
known and properly designed, the quantum mechanical motion is known and the
initial state (source state, $\left\vert \psi_{s}\right\rangle $) at $t=0$ can
potentially evolve to a given final state (target state, $\left\vert \psi
_{w}\right\rangle $) at $t=T$. In particular, for any instant $0\leq t\leq T$,
the probability $\mathcal{P}_{\left\vert \psi\left(  t\right)  \right\rangle
\rightarrow\left\vert \psi_{w}\right\rangle }$ that the system transitions
from the state $\left\vert \psi\left(  t\right)  \right\rangle $ to the state
$\left\vert \psi_{w}\right\rangle $ under the working assumption of constant
Hamiltonian is given by,%
\begin{equation}
\mathcal{P}_{\left\vert \psi\left(  t\right)  \right\rangle \rightarrow
\left\vert \psi_{w}\right\rangle }\overset{\text{def}}{=}\left\vert
\left\langle \psi_{w}|\psi\left(  t\right)  \right\rangle \right\vert
^{2}=\left\vert \left\langle \psi_{w}|e^{-\frac{i}{\hslash}\mathcal{H}t}%
|\psi_{s}\right\rangle \right\vert ^{2}\text{.}%
\end{equation}
The unitary operator $\mathcal{U}\left(  t\right)  \overset{\text{def}}%
{=}e^{-\frac{i}{\hslash}\mathcal{H}t}$ denotes the temporal evolution
operator. Fig. $1$ displays a graphical depiction of the digital (discrete
time) and analog (continuous time) quantum search algorithms.

Working in a continuous time quantum computing framework, Farhi and Gutmann
proposed an analog version of Grover's algorithm in Ref. \cite{farhi98} where
the state of the quantum register evolves continuously in time under the
action of a suitably chosen driving Hamiltonian (analog quantum computation).
Specifically, the transition probability from the source state $\left\vert
\psi_{s}\right\rangle $ to the target state $\left\vert \psi_{w}\right\rangle
$ after the application of the unitary continuous time evolution operator
$\mathcal{U}_{\text{FG}}\left(  t\right)  \overset{\text{def}}{=}e^{-\frac
{i}{\hslash}\mathcal{H}_{\text{FG}}t}$ for a closed quantum system described
by a constant Hamiltonian $\mathcal{H}_{\text{FG}}$ is given by,
\begin{equation}
\mathcal{P}_{\text{Farhi-Gutmann}}\left(  t\text{, }x\right)  \overset
{\text{def}}{=}\left\vert \left\langle \psi_{w}|e^{-\frac{i}{\hslash
}\mathcal{H}_{\text{FG}}t}|\psi_{s}\right\rangle \right\vert ^{2}=\sin
^{2}\left(  \frac{Ex}{\hslash}t\right)  +x^{2}\cos^{2}\left(  \frac
{Ex}{\hslash}t\right)  \text{,} \label{PFG}%
\end{equation}
where $E$ is a energy-like positive and \emph{real }constant coefficient. We
point out that $\mathcal{P}_{\text{Farhi-Gutmann}}$ in\ Eq. (\ref{PFG})
approaches one if $t$ approaches $h/(4Ex)$. For recent discussions on the
transition from the digital to analog quantum computational setting for
Grover's algorithm, we refer to Ref. \cite{carlo1,carlo2, cafaro2017}.

Ideally, one seeks to achieve unit success probability (that is, unit
fidelity) in the shortest possible time in a quantum search problem. There are
however, both practical and foundational issues that can justify the
exploration of alternative circumstances. For instance, from a practical
standpoint, one would desire to terminate a quantum information processing
task in the minimum possible time so as to mitigate decoherent effects that
can appear while controlling (by means of an external magnetic field, for
instance) the dynamics of a source state driven towards a target state
\cite{rabitz12,rabitz15,cappellaro18}. In addition, from a theoretical
viewpoint, it is known that no quantum measurement can perfectly discriminate
between two nonorthogonal pure states \cite{chefles00,croke09}. Moreover, it
is equally notorious that suitably engineered quantum measurements can enhance
the transition probability between two pure states \cite{fritz10}. Therefore,
minimizing the search time can be important from an experimental standpoint
while seeking at any cost perfect overlap between the final state and the
target state can be unnecessary from a purely foundational standpoint. Similar
lines of reasoning have paved the way to the fascinating exploration of a
possible tradeoff between fidelity and time optimal control of quantum unitary
transformations in Ref. \cite{rabitz12}.

In this paper, motivated by these issues and starting from the consideration
of a family of multi-parameter generalized quantum search Hamiltonians
originally introduced by Bae and Kwon in Ref. \cite{bae02}, we present a
detailed analysis concerning minimum search times and maximal success
probabilities that can be obtained from suitably chosen sub-families belonging
to the original family of Hamiltonians. In particular, we uncover the
existence of quantum search Hamiltonians characterized by minimum search times
needed for a perfect search that are smaller than the one required by the
Farhi-Gutmann perfect quantum search Hamiltonian. Furthermore, and more
interestingly, we report on the existence of imperfect quantum search
Hamiltonians that, despite being incapable of guaranteeing perfect search, can
outperform (in terms of minimum search time) perfect search Hamiltonians
provided that only a very large nearly optimal fidelity value is required to
stop the search.

The layout of the rest of the paper can be described as follows. In Section
II, we provide a detailed computation of a general expression for the
transition probability in the case of a quantum mechanical evolution governed
by a time-independent generalized quantum search Hamiltonian. In Section III,
we discuss a variety of limiting cases that arise from the generalized search
Hamiltonian. In particular, we distinguish optimal scenarios (that is, cases
where the probability of finding the target state equals one) from suboptimal
scenarios (that is, cases where the probability of finding the target state is
less than one). Our concluding remarks appear in Section IV. Finally,
technical details are presented in Appendices A, B, and C.

\begin{figure}[t]
\centering
\includegraphics[width=0.5\textwidth] {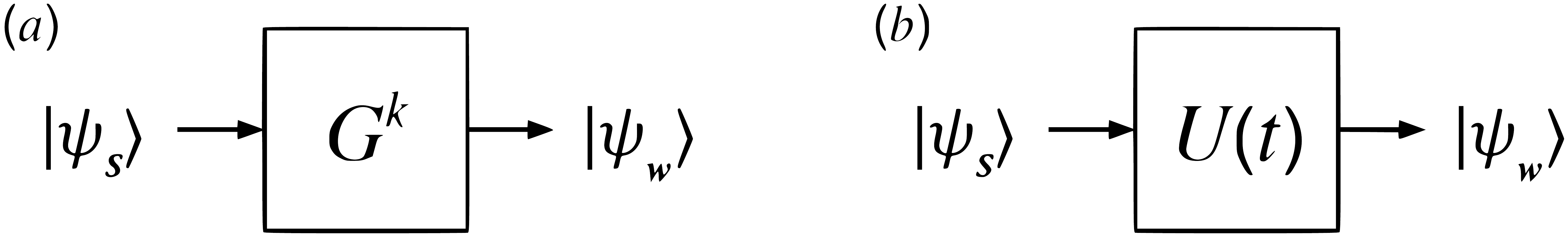}\caption{Gate-level schematic of
the (a) digital and (b) analog quantum search algorithms.}%
\label{fig1}%
\end{figure}

\section{Transition probability}

In this section, we consider the time-independent search Hamiltonian
$\mathcal{H}$ defined as \cite{bae02},%
\begin{equation}
\mathcal{H}\overset{\text{def}}{=}E\left[  \alpha\left\vert \psi
_{w}\right\rangle \left\langle \psi_{w}\right\vert +\beta\left\vert \psi
_{w}\right\rangle \left\langle \psi_{s}\right\vert +\gamma\left\vert \psi
_{s}\right\rangle \left\langle \psi_{w}\right\vert +\delta\left\vert \psi
_{s}\right\rangle \left\langle \psi_{s}\right\vert \right]  \text{.}
\label{hamilton}%
\end{equation}
The adimensional coefficients $\alpha$, $\beta$, $\gamma$, $\delta$ in Eq.
(\ref{hamilton}) are \emph{complex} expansion coefficients while $E$ is a
\emph{real} constant with energy physical dimensions. We also assume that the
quantum state $\left\vert \psi_{w}\right\rangle $ is the normalized target
state while $\left\vert \psi_{s}\right\rangle $ is the normalized initial
state with quantum overlap $x\overset{\text{def}}{=}\left\langle \psi_{w}%
|\psi_{s}\right\rangle $ that evolves unitarily according to the
Schr\"{o}dinger quantum mechanical evolution law,%
\begin{equation}
\left\vert \psi_{s}\right\rangle \mapsto e^{-\frac{i}{\hslash}\mathcal{H}%
t}\left\vert \psi_{s}\right\rangle \text{.}%
\end{equation}
In general, $x$ is a complex quantity. However, since any phase factor
$e^{i\phi_{ws}}$ with $\phi_{ws}\in%
%TCIMACRO{\U{211d} }%
%BeginExpansion
\mathbb{R}
%EndExpansion
$ in $x\overset{\text{def}}{=}\left\langle \psi_{w}|\psi_{s}\right\rangle
=\left\vert \left\langle \psi_{w}|\psi_{s}\right\rangle \right\vert
e^{i\phi_{ws}}$ can be incorporated into the state $\left\vert s\right\rangle
$, one can assume that $x\in%
%TCIMACRO{\U{211d} }%
%BeginExpansion
\mathbb{R}
%EndExpansion
_{+}\backslash\left\{  0\right\}  $. Our objective is to find the time
$t^{\ast}$ such that $\mathcal{P}\left(  t^{\ast}\right)  =\mathcal{P}_{\max}$
where $\mathcal{P}\left(  t\right)  $ is the transition probability defined as
\cite{sakurai,picasso},%
\begin{equation}
\mathcal{P}\left(  t\right)  \overset{\text{def}}{=}\left\vert \left\langle
\psi_{w}|e^{-\frac{i}{\hslash}\mathcal{H}t}|\psi_{s}\right\rangle \right\vert
^{2}\text{.} \label{fidelity}%
\end{equation}
\begin{figure}[t]
\centering
\includegraphics[width=0.3\textwidth] {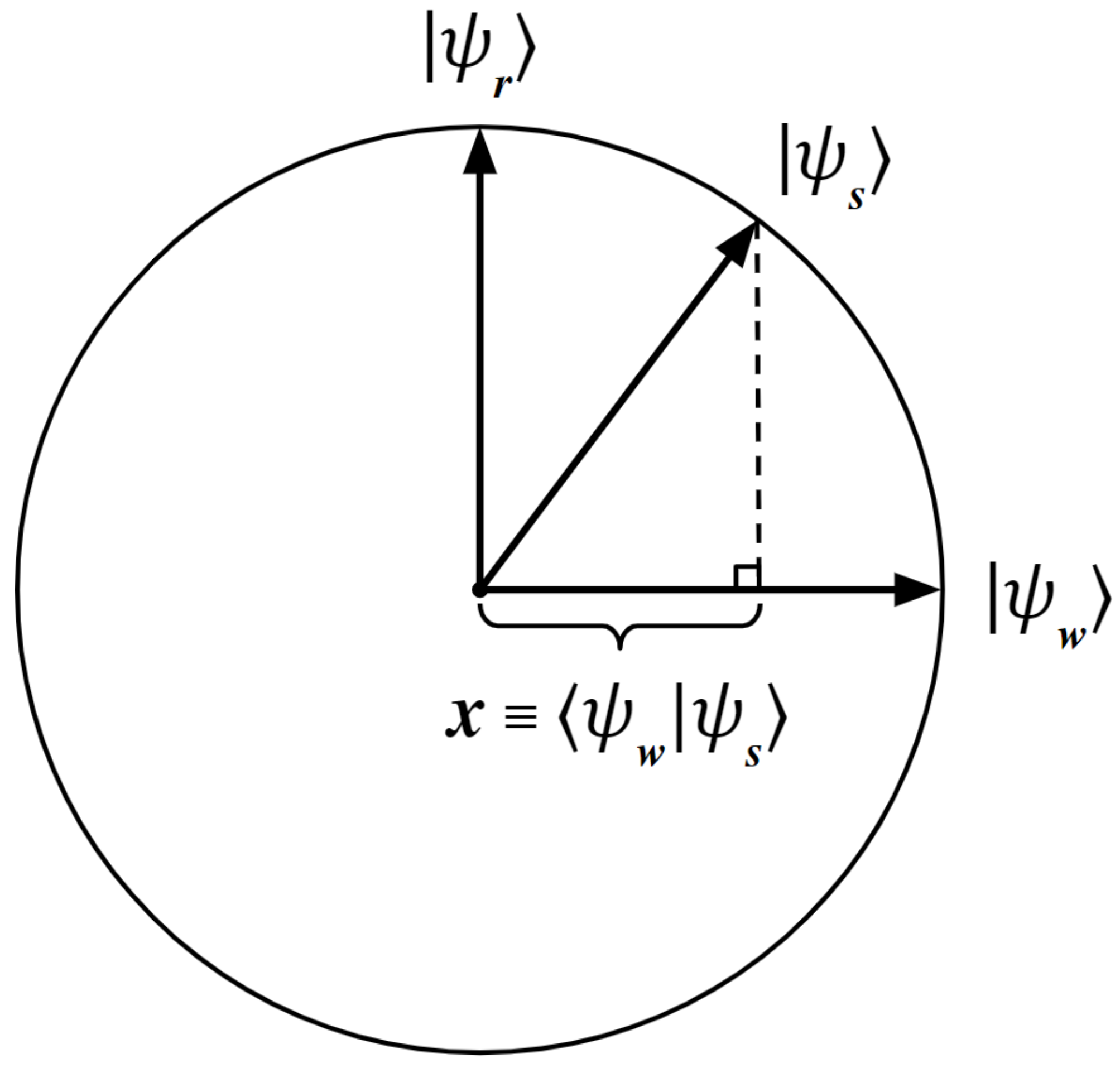}\caption{Visualization of the
normalized states $\left\vert \psi_{w}\right\rangle $, $\left\vert \psi
_{s}\right\rangle $, and $\left\vert \psi_{r}\right\rangle $, as well as the
quantum overlap $x$.}%
\label{fig2}%
\end{figure}Using the Gram-Schmidt orthonormalization technique, we can
construct an orthonormal set of quantum state vectors starting from the set
$\left\{  \left\vert \psi_{w}\right\rangle \text{, }\left\vert \psi
_{s}\right\rangle \right\}  $. The transition from a set of linear independent
state vectors to a set of orthonormal state vector can be described as,%
\begin{equation}
\left\{  \left\vert \psi_{w}\right\rangle \text{, }\left\vert \psi
_{s}\right\rangle \right\}  \rightarrow\left\{  \left\vert \psi_{w}%
\right\rangle \text{, }\left\vert \psi_{s}\right\rangle -\left\langle \psi
_{s}|\psi_{w}\right\rangle \left\vert \psi_{w}\right\rangle \right\}
\rightarrow\left\{  \frac{\left\vert \psi_{w}\right\rangle }{\left\Vert
\left\vert \psi_{w}\right\rangle \right\Vert }\text{, }\frac{\left\vert
\psi_{s}\right\rangle -\left\langle \psi_{s}|\psi_{w}\right\rangle \left\vert
\psi_{w}\right\rangle }{\left\Vert \left\vert \psi_{s}\right\rangle
-\left\langle \psi_{s}|\psi_{w}\right\rangle \left\vert \psi_{w}\right\rangle
\right\Vert }\right\}  \text{.}%
\end{equation}
For notational simplicity, let us define the quantum state vector $\left\vert
\psi_{r}\right\rangle $ as%
\begin{equation}
\left\vert \psi_{r}\right\rangle \overset{\text{def}}{=}\frac{\left\vert
\psi_{s}\right\rangle -\left\langle \psi_{s}|\psi_{w}\right\rangle \left\vert
\psi_{w}\right\rangle }{\left\Vert \left\vert \psi_{s}\right\rangle
-\left\langle \psi_{s}|\psi_{w}\right\rangle \left\vert \psi_{w}\right\rangle
\right\Vert }\text{.} \label{erre2}%
\end{equation}
Recalling the definition of the quantum overlap $x$, Eq. (\ref{erre2}) can be
expressed as,%
\begin{equation}
\left\vert \psi_{r}\right\rangle =\frac{\left\vert \psi_{s}\right\rangle
-\left\langle \psi_{s}|\psi_{w}\right\rangle \left\vert \psi_{w}\right\rangle
}{\sqrt{\left\langle \psi_{s}|\psi_{s}\right\rangle -\left\langle \psi
_{s}|\psi_{w}\right\rangle ^{2}}}=\frac{1}{\sqrt{1-x^{2}}}\left(  \left\vert
\psi_{s}\right\rangle -x\left\vert \psi_{w}\right\rangle \right)  \text{.}
\label{fiar}%
\end{equation}
Fig. $2$ displays a graphical depiction of the orthonormal states $\left\{
\left\vert \psi_{w}\right\rangle \text{, }\left\vert \psi_{r}\right\rangle
\right\}  $ together with the source state $\left\vert \psi_{s}\right\rangle $
and the quantum overlap $x$. Fig. $3$, instead, is a simple depiction of the
orthogonalization and normalization procedures that specify the Gram-Schmidt
orthonormalization technique. Note that because of the definition of
$\left\vert \psi_{r}\right\rangle $ in Eq. (\ref{fiar}), $x$ must be different
from one. In terms of the set of orthonormal basis vectors $\left\{
\left\vert \psi_{w}\right\rangle \text{, }\left\vert \psi_{r}\right\rangle
\right\}  $, the source state $\left\vert \psi_{s}\right\rangle $ becomes%
\begin{equation}
\left\vert \psi_{s}\right\rangle =\left\vert \psi_{s}\right\rangle \left(
\left\vert \psi_{w}\right\rangle \left\langle \psi_{w}\right\vert +\left\vert
\psi_{r}\right\rangle \left\langle \psi_{r}\right\vert \right)  =\left\langle
\psi_{w}|\psi_{s}\right\rangle \left\vert \psi_{w}\right\rangle +\left\langle
\psi_{r}|\psi_{s}\right\rangle \left\vert \psi_{r}\right\rangle \text{.}
\label{chell}%
\end{equation}
Note that the quantum mechanical overlap $\left\langle \psi_{r}|\psi
_{s}\right\rangle $ in Eq. (\ref{chell}) can be recast as,
\begin{equation}
\left\langle \psi_{r}|\psi_{s}\right\rangle =\frac{1}{\sqrt{1-x^{2}}}\left(
\left\langle \psi_{s}\right\vert -x\left\langle \psi_{w}\right\vert \right)
\left(  \left\vert \psi_{s}\right\rangle \right)  =\frac{1}{\sqrt{1-x^{2}}%
}\left(  1-x^{2}\right)  =\sqrt{1-x^{2}}\text{.} \label{chist}%
\end{equation}
Therefore, by\textbf{ }using Eq. (\ref{chist}), the state $\left\vert \psi
_{s}\right\rangle $ in Eq. (\ref{chell}) becomes
\begin{equation}
\left\vert \psi_{s}\right\rangle =x\left\vert \psi_{w}\right\rangle
+\sqrt{1-x^{2}}\left\vert \psi_{r}\right\rangle \text{.} \label{sr}%
\end{equation}
The matrix representation of the Hamiltonian $\mathcal{H}$ in Eq.
(\ref{hamilton}) with respect to the orthonormal basis $\left\{  \left\vert
\psi_{w}\right\rangle \text{, }\left\vert \psi_{r}\right\rangle \right\}  $
where $\left\langle \psi_{w}|\psi_{r}\right\rangle =\delta_{wr}$, with
$\delta_{wr}$ denoting the Kronecker delta, can be formally written as%
\begin{equation}
\left[  \mathcal{H}\right]  _{\left\{  \left\vert \psi_{w}\right\rangle
\text{, }\left\vert \psi_{r}\right\rangle \right\}  }\overset{\text{def}}%
{=}\left(
\begin{array}
[c]{cc}%
\left\langle \psi_{w}|\mathcal{H}|\psi_{w}\right\rangle  & \left\langle
\psi_{w}|\mathcal{H}|\psi_{r}\right\rangle \\
\left\langle \psi_{r}|\mathcal{H}|\psi_{w}\right\rangle  & \left\langle
\psi_{r}|\mathcal{H}|\psi_{r}\right\rangle
\end{array}
\right)  \text{.}%
\end{equation}
Using Eqs. (\ref{hamilton}) and (\ref{sr}) together with the orthonormality
conditions $\left\langle \psi_{w}|\psi_{r}\right\rangle =\delta_{wr}$, we have%
\begin{equation}
\left[  \mathcal{H}\right]  _{\left\{  \left\vert \psi_{w}\right\rangle
\text{, }\left\vert \psi_{r}\right\rangle \right\}  }=\left(
\begin{array}
[c]{cc}%
H_{11} & H_{12}\\
H_{21} & H_{22}%
\end{array}
\right)  \text{,} \label{symm}%
\end{equation}
where,%
\begin{align}
&  H_{11}\overset{\text{def}}{=}E\left[  \alpha+\left(  \beta+\gamma\right)
x+\delta x^{2}\right]  \text{, }H_{12}\overset{\text{def}}{=}E\sqrt{1-x^{2}%
}\left(  \beta+\delta x\right)  \text{,}\nonumber\\
& \nonumber\\
&  H_{21}\overset{\text{def}}{=}E\sqrt{1-x^{2}}\left(  \gamma+\delta x\right)
\text{, }H_{22}\overset{\text{def}}{=}E\delta\left(  1-x^{2}\right)  \text{.}
\label{heq}%
\end{align}
\begin{figure}[t]
\centering
\includegraphics[width=0.75\textwidth] {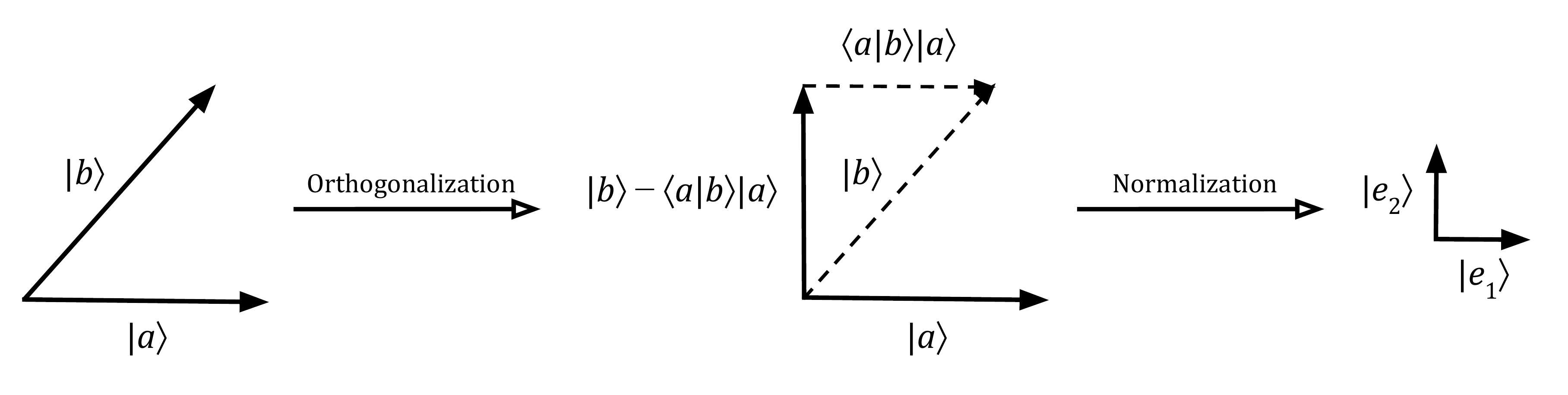}\caption{Illustration of the
Gram-Schmidt orthonormalization procedure for some vectors $\left\vert
a\right\rangle $ and $\left\vert b\right\rangle $. In this case, the resulting
orthonormal basis consists of $\left\vert e_{1}\right\rangle \overset
{\text{def}}{=}\frac{\left\vert a\right\rangle }{\left\Vert \left\vert
a\right\rangle \right\Vert }$ and $\left\vert e_{2}\right\rangle
\overset{\text{def}}{=}\frac{\left\vert b\right\rangle -\left\langle
a|b\right\rangle \left\vert a\right\rangle }{\left\Vert \left\vert
b\right\rangle -\left\langle a|b\right\rangle \left\vert a\right\rangle
\right\Vert }$.}%
\label{fig3}%
\end{figure}Observe that the Hamiltonian $\mathcal{H}$ is an Hermitian
operator and, therefore, its eigenvalues must be \emph{real }(for details, see
Appendix A). For this reason, recalling the previous constraints on $x$, we
finally have $x\in\left(  0\text{,}1\right)  $. Furthermore, imposing that
$\mathcal{H}=\mathcal{H}^{\dagger}$ where the dagger symbol \textquotedblleft%
$\dagger$\textquotedblright\ is the Hermitian conjugation operation, we have
\begin{equation}
\left(
\begin{array}
[c]{cc}%
H_{11} & H_{12}\\
H_{21} & H_{22}%
\end{array}
\right)  =\left(
\begin{array}
[c]{cc}%
H_{11}^{\ast} & H_{21}^{\ast}\\
H_{12}^{\ast} & H_{22}^{\ast}%
\end{array}
\right)  \text{.} \label{heq2}%
\end{equation}
Then, from Eqs. (\ref{heq}) and (\ref{heq2}), it follows that $\alpha$ and
$\delta$ must be \emph{real} coefficients while $\beta=\gamma^{\ast}$. The
symbol \textquotedblleft$\ast$\textquotedblright\ denotes the \emph{complex}
conjugation operation. Next, let us diagonalize the Hermitian matrix $\left[
\mathcal{H}\right]  _{\left\{  \left\vert \psi_{w}\right\rangle \text{,
}\left\vert \psi_{r}\right\rangle \right\}  }$ in Eq. (\ref{symm}). The two
\emph{real} eigenvalues $\lambda_{\pm}$ of the matrix can be written as,
\begin{equation}
\lambda_{\pm}\overset{\text{def}}{=}\frac{1}{2}\left[  \left(  H_{11}%
+H_{22}\right)  \pm\sqrt{\left(  H_{11}-H_{22}\right)  ^{2}+4H_{12}H_{21}%
}\right]  \text{.} \label{eigen}%
\end{equation}
The eigenspaces $\mathcal{E}_{\lambda_{-}}$ and $\mathcal{E}_{\lambda_{+}}$
that correspond to the eigenvalues $\lambda_{-}$ and $\lambda_{+}$ are defined
as%
\begin{equation}
\mathcal{E}_{\lambda_{-}}\overset{\text{def}}{=}\text{\textrm{Span}}\left\{
\left\vert v_{\lambda_{-}}\right\rangle \right\}  \text{, and }\mathcal{E}%
_{\lambda_{+}}\overset{\text{def}}{=}\text{\textrm{Span}}\left\{  \left\vert
v_{\lambda_{+}}\right\rangle \right\}  \text{,}%
\end{equation}
respectively. Furthermore, two orthogonal eigenvectors $\left\vert
v_{\lambda_{+}}\right\rangle $ and $\left\vert v_{\lambda_{-}}\right\rangle $
corresponding to the distinct eigenvalues $\lambda_{+}$ and $\lambda_{-}$ are
given by,%
\begin{equation}
\left\vert v_{\lambda_{+}}\right\rangle \overset{\text{def}}{=}\left(
\begin{array}
[c]{c}%
\frac{1}{2H_{21}}\left[  \left(  H_{11}-H_{22}\right)  +\sqrt{\left(
H_{11}-H_{22}\right)  ^{2}+4H_{12}H_{21}}\right] \\
1
\end{array}
\right)  \text{,} \label{v1}%
\end{equation}
and,%
\begin{equation}
\left\vert v_{\lambda_{-}}\right\rangle \overset{\text{def}}{=}\left(
\begin{array}
[c]{c}%
\frac{1}{2H_{21}}\left[  \left(  H_{11}-H_{22}\right)  -\sqrt{\left(
H_{11}-H_{22}\right)  ^{2}+4H_{12}H_{21}}\right] \\
1
\end{array}
\right)  \text{,} \label{v2}%
\end{equation}
respectively. For notational simplicity, let us introduce two \emph{complex}
quantities $\mathcal{A}$ and $\mathcal{B}$ defined as%
\begin{equation}
\mathcal{A}\overset{\text{def}}{=}\frac{1}{2H_{21}}\left[  \left(
H_{11}-H_{22}\right)  -\sqrt{\left(  H_{11}-H_{22}\right)  ^{2}+4H_{12}H_{21}%
}\right]  \text{,} \label{anew}%
\end{equation}
and,%
\begin{equation}
\mathcal{B}\overset{\text{def}}{=}\frac{1}{2H_{21}}\left[  \left(
H_{11}-H_{22}\right)  +\sqrt{\left(  H_{11}-H_{22}\right)  ^{2}+4H_{12}H_{21}%
}\right]  \text{,} \label{bnew}%
\end{equation}
respectively. Using Eqs. (\ref{v1}), (\ref{v2}), (\ref{anew}), and
(\ref{bnew}), the eigenvector matrix $M_{\mathcal{H}}$ for the matrix $\left[
\mathcal{H}\right]  _{\left\{  \left\vert \psi_{w}\right\rangle \text{,
}\left\vert \psi_{r}\right\rangle \right\}  }$ and its inverse $M_{\mathcal{H}%
}^{-1}$ can be formally written as,%
\begin{equation}
M_{\mathcal{H}}\overset{\text{def}}{=}\left(
\begin{array}
[c]{cc}%
\mathcal{A} & \mathcal{B}\\
1 & 1
\end{array}
\right)  \text{,} \label{mmatrix2}%
\end{equation}
and,%
\begin{equation}
M_{\mathcal{H}}^{-1}\overset{\text{def}}{=}\frac{1}{\mathcal{A}-\mathcal{B}%
}\left(
\begin{array}
[c]{cc}%
1 & -\mathcal{B}\\
-1 & \mathcal{A}%
\end{array}
\right)  =M_{\mathcal{H}}^{\dagger}\text{,} \label{mimatrix2}%
\end{equation}
respectively. In terms of the matrices $M_{\mathcal{H}}$ in Eq.
(\ref{mmatrix2}), $M_{\mathcal{H}}^{-1}$ in Eq. (\ref{mimatrix2}), and the
diagonal matrix $H_{\text{diagonal}}$ defined as,%
\begin{equation}
H_{\text{diagonal}}\overset{\text{def}}{=}\left[  \mathcal{H}\right]
_{\left\{  \left\vert v_{\lambda-}\right\rangle \text{, }\left\vert
v_{\lambda_{+}}\right\rangle \right\}  }=\left(
\begin{array}
[c]{cc}%
\left\langle v_{\lambda-}|\mathcal{H}|v_{\lambda-}\right\rangle  &
\left\langle v_{\lambda-}|\mathcal{H}|v_{\lambda_{+}}\right\rangle \\
\left\langle v_{\lambda_{+}}|\mathcal{H}|v_{\lambda-}\right\rangle  &
\left\langle v_{\lambda_{+}}|\mathcal{H}|v_{\lambda_{+}}\right\rangle
\end{array}
\right)  =\left(
\begin{array}
[c]{cc}%
\lambda_{-} & 0\\
0 & \lambda_{+}%
\end{array}
\right)  \text{,} \label{hdiagonal}%
\end{equation}
the matrix $\left[  \mathcal{H}\right]  _{\left\{  \left\vert \psi
_{w}\right\rangle \text{, }\left\vert \psi_{r}\right\rangle \right\}  }$
in\ Eq. (\ref{symm}) becomes%
\begin{equation}
\left[  \mathcal{H}\right]  _{\left\{  \left\vert \psi_{w}\right\rangle
\text{, }\left\vert \psi_{r}\right\rangle \right\}  }=M_{\mathcal{H}%
}H_{\text{diagonal}}M_{\mathcal{H}}^{-1}=\left(
\begin{array}
[c]{cc}%
\mathcal{A} & \mathcal{B}\\
1 & 1
\end{array}
\right)  \left(
\begin{array}
[c]{cc}%
\lambda_{-} & 0\\
0 & \lambda_{+}%
\end{array}
\right)  \left(
\begin{array}
[c]{cc}%
\frac{1}{\mathcal{A}-\mathcal{B}} & \frac{-\mathcal{B}}{\mathcal{A}%
-\mathcal{B}}\\
\frac{-1}{\mathcal{A}-\mathcal{B}} & \frac{\mathcal{A}}{\mathcal{A}%
-\mathcal{B}}%
\end{array}
\right)  \text{.}%
\end{equation}
We recall that the eigenvalues in\ Eq. (\ref{hdiagonal}) are defined in\ Eq.
(\ref{eigen}) while $\mathcal{A}$ and $\mathcal{B}$ appear in\ Eqs.
(\ref{anew}) and (\ref{bnew}), respectively. At this juncture, we also recall
that our objective is to find the time $t^{\ast}$ such that $\mathcal{P}%
\left(  t^{\ast}\right)  =\mathcal{P}_{\max}$ where the transition probability
$\mathcal{P}\left(  t\right)  $ is defined in Eq. (\ref{fidelity}). Employing
standard linear algebra techniques, $\mathcal{P}\left(  t\right)  $ can be
recast as%
\begin{equation}
\mathcal{P}\left(  t\right)  \overset{\text{def}}{=}\left\vert \left\langle
\psi_{w}|e^{-\frac{i}{\hslash}\mathcal{H}t}|\psi_{s}\right\rangle \right\vert
^{2}=\left\vert \left\langle \psi_{w}|e^{-\frac{i}{\hslash}M\mathcal{H}%
_{\text{diagonal}}M^{\dagger}t}|\psi_{s}\right\rangle \right\vert
^{2}=\left\vert \left\langle \psi_{w}|Me^{-\frac{i}{\hslash}\mathcal{H}%
_{\text{diagonal}}t}M^{\dagger}|\psi_{s}\right\rangle \right\vert ^{2}\text{,}
\label{pt3}%
\end{equation}
where $\mathcal{H}_{\text{diagonal}}$ denotes the Hermitian operator whose
matrix representation is $H_{\text{diagonal}}$ in Eq. (\ref{hdiagonal}). Using
the matrix notation with components expressed with respect to the orthonormal
basis $\left\{  \left\vert \psi_{w}\right\rangle \text{, }\left\vert \psi
_{r}\right\rangle \right\}  $, quantum states $\left\vert \psi_{w}%
\right\rangle $ and $\left\vert \psi_{s}\right\rangle $ are given by
\begin{equation}
\left\vert \psi_{w}\right\rangle \overset{\text{def}}{=}\left(
\begin{array}
[c]{c}%
1\\
0
\end{array}
\right)  \text{, and }\left\vert \psi_{s}\right\rangle \overset{\text{def}}%
{=}\left(
\begin{array}
[c]{c}%
x\\
\sqrt{1-x^{2}}%
\end{array}
\right)  \text{,} \label{matic2}%
\end{equation}
respectively. By means of Eqs. (\ref{mmatrix2}), (\ref{mimatrix2}), and
(\ref{matic2}), the quantum state amplitude $\left\langle \psi_{w}%
|e^{-\frac{i}{\hslash}\mathcal{H}t}|\psi_{s}\right\rangle $ that appears in
the expression of the fidelity $\mathcal{P}\left(  t\right)  $ in\ Eq.
(\ref{pt3}) becomes%
\begin{equation}
\left\langle \psi_{w}|e^{-\frac{i}{\hslash}\mathcal{H}t}|\psi_{s}\right\rangle
=\frac{1}{\mathcal{A}-\mathcal{B}}\left[  \mathcal{A}e^{-\frac{i}{\hslash
}\lambda_{-}t}\left(  x-\mathcal{B}\sqrt{1-x^{2}}\right)  -\mathcal{B}%
e^{-\frac{i}{\hslash}\lambda_{+}t}\left(  x-\mathcal{A}\sqrt{1-x^{2}}\right)
\right]  \text{,} \label{part1b}%
\end{equation}
and, as a consequence, its complex conjugate $\left\langle \psi_{w}%
|e^{-\frac{i}{\hslash}\mathcal{H}t}|\psi_{s}\right\rangle ^{\ast}$ is,
\begin{equation}
\left\langle \psi_{w}|e^{-\frac{i}{\hslash}\mathcal{H}t}|\psi_{s}\right\rangle
^{\ast}=\frac{1}{\mathcal{A}-\mathcal{B}}\left[  \mathcal{A}e^{\frac
{i}{\hslash}\lambda_{-}t}\left(  x-\mathcal{B}\sqrt{1-x^{2}}\right)
-\mathcal{B}e^{\frac{i}{\hslash}\lambda_{+}t}\left(  x-\mathcal{A}%
\sqrt{1-x^{2}}\right)  \right]  \text{.} \label{part2b}%
\end{equation}
Observe that,%
\begin{equation}
e^{-\frac{i}{\hslash}\lambda_{-}t}=e^{-\frac{i}{\hslash}\frac{H_{11}+H_{22}%
}{2}t}e^{i\frac{\mathrm{a}}{\hslash}t}\text{ and, }e^{-\frac{i}{\hslash
}\lambda_{+}t}=e^{-\frac{i}{\hslash}\frac{H_{11}+H_{22}}{2}t}e^{-i\frac
{\mathrm{a}}{\hslash}t} \label{aeq}%
\end{equation}
where, recalling Eq. (\ref{eigen}), the \emph{real} quantity $\mathrm{a}$ is
defined as%
\begin{equation}
\mathrm{a}\overset{\text{def}}{=}\frac{1}{2}\sqrt{\left(  H_{11}%
-H_{22}\right)  ^{2}+4H_{12}H_{21}}\text{.} \label{adef}%
\end{equation}
Employing Eq. (\ref{aeq}), the \emph{complex} probability amplitudes in Eqs.
(\ref{part1b}) and (\ref{part2b}) become%
\begin{equation}
\left\langle \psi_{w}|e^{-\frac{i}{\hslash}\mathcal{H}t}|\psi_{s}\right\rangle
=e^{-\frac{i}{\hslash}\frac{H_{11}+H_{22}}{2}t}\left[  \frac{\mathcal{A}%
\left(  x-\mathcal{B}\sqrt{1-x^{2}}\right)  }{\mathcal{A}-\mathcal{B}%
}e^{i\frac{\mathrm{a}}{\hslash}t}-\frac{\mathcal{B}\left(  x-\mathcal{A}%
\sqrt{1-x^{2}}\right)  }{\mathcal{A}-\mathcal{B}}e^{-i\frac{\mathrm{a}%
}{\hslash}t}\right]  \text{,} \label{part3}%
\end{equation}
and,%
\begin{equation}
\left\langle \psi_{w}|e^{-\frac{i}{\hslash}\mathcal{H}t}|\psi_{s}\right\rangle
^{\ast}=e^{\frac{i}{\hslash}\frac{H_{11}+H_{22}}{2}t}\left[  \frac
{\mathcal{A}^{\ast}\left(  x-\mathcal{B}^{\ast}\sqrt{1-x^{2}}\right)
}{\mathcal{A}^{\ast}-\mathcal{B}^{\ast}}e^{-i\frac{\mathrm{a}}{\hslash}%
t}-\frac{\mathcal{B}^{\ast}\left(  x-\mathcal{A}^{\ast}\sqrt{1-x^{2}}\right)
}{\mathcal{A}^{\ast}-\mathcal{B}^{\ast}}e^{i\frac{\mathrm{a}}{\hslash}%
t}\right]  \text{,} \label{part4}%
\end{equation}
respectively. Using Eqs. (\ref{part3})\ and (\ref{part4}) and introducing the
following three quantities%
\begin{equation}
\tilde{A}\overset{\text{def}}{=}\frac{\mathcal{A}\left(  x-\mathcal{B}%
\sqrt{1-x^{2}}\right)  }{\mathcal{A}-\mathcal{B}}\text{, }\tilde{B}%
\overset{\text{def}}{=}-\frac{\mathcal{B}\left(  x-\mathcal{A}\sqrt{1-x^{2}%
}\right)  }{\mathcal{A}-\mathcal{B}}\text{, and }\tilde{\alpha}=\frac
{\mathrm{a}}{\hslash}t\text{,} \label{newroba}%
\end{equation}
the transition probability $\mathcal{P}\left(  t\right)  $ in Eq.
(\ref{fidelity}) can be written as%
\begin{equation}
\mathcal{P}\left(  \tilde{\alpha}\right)  =\left[  \tilde{A}e^{i\tilde{\alpha
}}+\tilde{B}e^{-i\tilde{\alpha}}\right]  \left[  \tilde{A}^{\ast}%
e^{-i\tilde{\alpha}}+\tilde{B}^{\ast}e^{i\tilde{\alpha}}\right]  =\left\vert
\tilde{A}\right\vert ^{2}+\left\vert \tilde{B}\right\vert ^{2}+2\tilde
{A}\tilde{B}^{\ast}\cos\left(  2\tilde{\alpha}\right)  \text{,}%
\end{equation}
where we point out that $\tilde{A}\tilde{B}^{\ast}$ is \emph{real} since
$H_{12}=H_{21}^{\ast}$. By employing standard trigonometric identities in a
convenient sequential order (for details, see Appendix B), we find%
\begin{equation}
\mathcal{P}\left(  \tilde{\alpha}\right)  =\left\vert \tilde{A}-\tilde
{B}\right\vert ^{2}\sin^{2}\left(  \tilde{\alpha}\right)  +\left\vert
\tilde{A}+\tilde{B}\right\vert ^{2}\cos^{2}\left(  \tilde{\alpha}\right)
\text{.} \label{fess}%
\end{equation}
Using Eqs. (\ref{newroba}), (\ref{adef}), (\ref{anew}), and (\ref{bnew}), the
transition probability $\mathcal{P}\left(  \tilde{\alpha}\right)  $ in Eq.
(\ref{fess}) becomes%
\begin{equation}
\mathcal{P}\left(  t\right)  =\frac{\left\vert \left(  H_{11}-H_{22}\right)
x+2H_{12}\sqrt{1-x^{2}}\right\vert ^{2}}{\left(  H_{11}-H_{22}\right)
^{2}+4H_{12}H_{21}}\sin^{2}\left(  \frac{\sqrt{\left(  H_{11}-H_{22}\right)
^{2}+4H_{12}H_{21}}}{2\hslash}t\right)  +x^{2}\cos^{2}\left(  \frac
{\sqrt{\left(  H_{11}-H_{22}\right)  ^{2}+4H_{12}H_{21}}}{2\hslash}t\right)
\text{.} \label{it}%
\end{equation}
From Eq. (\ref{it}), it follows that the maximum $\mathcal{P}_{\max
}=\mathcal{P}\left(  t^{\ast}\right)  $ of $\mathcal{P}\left(  t\right)  $
occurs at the instant $t^{\ast}$,%
\begin{equation}
t^{\ast}\overset{\text{def}}{=}\frac{\pi\hslash}{\sqrt{\left(  H_{11}%
-H_{22}\right)  ^{2}+4H_{12}H_{21}}}\text{,} \label{start}%
\end{equation}
and equals%
\begin{equation}
\mathcal{P}_{\max}=\frac{\left\vert \left(  H_{11}-H_{22}\right)
x+2H_{12}\sqrt{1-x^{2}}\right\vert ^{2}}{\left(  H_{11}-H_{22}\right)
^{2}+4H_{12}H_{21}}\text{.} \label{maxp}%
\end{equation}
Finally, making use of Eq.(\ref{heq}) and recalling that $\alpha$ and $\delta$
must be \emph{real} coefficients while $\beta=\gamma^{\ast}$, $\mathcal{P}%
_{\max}$ in Eq. (\ref{maxp}) becomes%
\begin{equation}
\mathcal{P}_{\max}\left(  \alpha\text{, }\beta\text{, }\delta\text{,
}x\right)  =\frac{\left\vert \left[  \left(  \alpha-\delta\right)  +\left(
\beta+\beta^{\ast}\right)  x+2\delta x^{2}\right]  x+2\left(  \beta+\delta
x\right)  \left(  1-x^{2}\right)  \right\vert ^{2}}{\left[  \left(
\alpha-\delta\right)  +\left(  \beta+\beta^{\ast}\right)  x+2\delta
x^{2}\right]  ^{2}+4\left(  1-x^{2}\right)  \left(  \beta+\delta x\right)
\left(  \beta^{\ast}+\delta x\right)  }\text{.} \label{nono}%
\end{equation}
Note that $\gamma=\beta^{\ast}$, $\beta+\beta^{\ast}=2\operatorname{Re}\left(
\beta\right)  \in%
%TCIMACRO{\U{211d} }%
%BeginExpansion
\mathbb{R}
%EndExpansion
$, and the product $\left(  \beta+\delta x\right)  \left(  \beta^{\ast}+\delta
x\right)  $ is a \emph{real} quantity for any \emph{complex} parameter $\beta$.

\section{Discussion}

In this section, we discuss a variety of limiting cases that arise from the
generalized search Hamiltonian in Eq. (\ref{hamilton}). In particular, we make
a distinction between optimal and suboptimal scenarios. The former scenarios
are cases where the probability of finding the target state equals one. The
latter scenarios, instead, are cases where the probability of finding the
target state is less than one.

\emph{General Case}: The general case is specified by the conditions
$\alpha\neq\delta$ \emph{real} and $\beta=\gamma^{\ast}$ \emph{complex
}coefficients. We note that after some straightforward but tedious algebra,
$\mathcal{P}_{\max}$ in Eq. (\ref{nono}) can be recast as%
\begin{equation}
\mathcal{P}_{\max}\left(  \alpha\text{, }\beta\text{, }\delta\text{,
}x\right)  =\frac{4\left[  \left\vert \beta\right\vert ^{2}-\operatorname{Re}%
^{2}\left(  \beta\right)  \right]  x^{4}+\left[  \left(  \alpha+\delta\right)
^{2}-8\left(  \left\vert \beta\right\vert ^{2}-\operatorname{Re}^{2}\left(
\beta\right)  \right)  \right]  x^{2}+4\operatorname{Re}\left(  \beta\right)
\left(  \alpha+\delta\right)  x+4\left\vert \beta\right\vert ^{2}}{4\left[
\alpha\delta+\operatorname{Re}^{2}\left(  \beta\right)  -\left\vert
\beta\right\vert ^{2}\right]  x^{2}+4\operatorname{Re}\left(  \beta\right)
\left(  \alpha+\delta\right)  x+\left[  \left(  \alpha-\delta\right)
^{2}+4\left\vert \beta\right\vert ^{2}\right]  }\text{.} \label{nano2}%
\end{equation}
Furthermore, by using Eq. (\ref{heq}) in Eq. (\ref{start}), the time $t^{\ast
}$ at which this maximum transition probability value $\mathcal{P}_{\max}$ is
achieved becomes,%
\begin{equation}
t_{\mathcal{H}}^{\ast}\overset{\text{def}}{=}\frac{\pi\hslash}{E\sqrt{\left[
\left(  \alpha-\delta\right)  +x\left(  \beta+\beta^{\ast}\right)
+2x^{2}\delta\right]  ^{2}+4\left(  1-x^{2}\right)  \left(  \beta+\delta
x\right)  \left(  \beta^{\ast}+\delta x\right)  }}\text{.} \label{nano}%
\end{equation}
The subscript $\mathcal{H}$ in $t_{\mathcal{H}}^{\ast}$ denotes the
generalized search Hamiltonian in Eq. (\ref{hamilton}). Observe that
$t_{\mathcal{H}}^{\ast}$ in Eq. (\ref{nano}) can be rewritten as,%
\begin{equation}
t_{\mathcal{H}}^{\ast}\overset{\text{def}}{=}\frac{2}{\sqrt{4\left[
\alpha\delta+\operatorname{Re}^{2}\left(  \beta\right)  -\left\vert
\beta\right\vert ^{2}\right]  x^{2}+4\operatorname{Re}\left(  \beta\right)
\left(  \alpha+\delta\right)  x+\left[  \left(  \alpha-\delta\right)
^{2}+4\left\vert \beta\right\vert ^{2}\right]  }}\frac{\pi\hslash}{2E}\text{.}
\label{nano1}%
\end{equation}
\begin{table}[t]
\centering
\begin{tabular}
[c]{c|c|c|c|c|c}%
Case & $\alpha$ & $\beta$ & $\gamma$ & $\delta$ & ${\mathcal{P}}_{\text{max}}%
$\\\hline
General & $\neq\delta$ & $\gamma^{*}\in\mathbb{C}$ & $\beta^{*}\in\mathbb{C}$
& $\neq\alpha$ & $\leq1$\\
1 & $\delta$ & $0$ & $0$ & $\alpha$ & $1$\\
2 & $\neq\delta$ & $0$ & $0$ & $\neq\alpha$ & $\leq1$\\
3 & $0$ & $\gamma^{*}\in\mathbb{R}$ & $\beta^{*}\in\mathbb{R}$ & $0$ & $1$\\
4 & $0$ & $\gamma*\in\mathbb{C}$ & $\beta^{*}\in\mathbb{C}$ & $0$ & $\leq1$\\
5 & $\delta$ & $\gamma^{*}\in\mathbb{R}$ & $\beta^{*}\in\mathbb{R}$ & $\alpha$
& $1$\\
6 & $\delta$ & $\gamma^{*}\in\mathbb{C}$ & $\beta^{*}\in\mathbb{C}$ & $\alpha$
& $\leq1$\\
7 & $\neq\delta$ & $\gamma^{*}\in\mathbb{R}$ & $\beta^{*}\in\mathbb{R}$ &
$\neq\alpha$ & $\leq1$%
\end{tabular}
\caption{Summary of maximal success probability values $\mathcal{P}_{\max}$
that can be achieved for a variety of choices of the parameters $\alpha$,
$\beta$, $\gamma$, and $\delta$ specifying the quantum search Hamiltonian
$\mathcal{H}$ in Eq. (\ref{hamilton}).}%
\end{table}In what follows, we choose to briefly discuss a number of limiting
cases that arise from Eq. (\ref{nano2}). In particular, the
big-calligraphic-$\mathcal{O}$ notation $f\left(  \varepsilon\right)
=\mathcal{O}\left(  g\left(  \varepsilon\right)  \right)  $ means that
$f\left(  \varepsilon\right)  $ is an infinitesimal of order equal to
$g\left(  \varepsilon\right)  $ as $\varepsilon$ approaches zero, that is,%
\begin{equation}
\lim_{\varepsilon\rightarrow0}\frac{f\left(  \varepsilon\right)  }{g\left(
\varepsilon\right)  }=K\text{,}%
\end{equation}
where $K$ denotes a nonzero \emph{real} constant. In Table I we report an
overview of the maximal success probability values that can be obtained for a
variety of choices of the parameters $\alpha$, $\beta$, $\gamma$, and $\delta$
characterizing the quantum search Hamiltonian $\mathcal{H}$. In particular, we
note that the unit success probability $\mathcal{P}_{\max}=1$ can be achieved
only when considering the Hamiltonians $\mathcal{H}_{1}$, $\mathcal{H}_{3}$,
and $\mathcal{H}_{5}$. Fig. $4$, instead, displays the negative effect on the
maximal success probability $\mathcal{P}_{\max}$ due to asymmetries
($\alpha\neq\delta$) and complexities ($\beta\in%
%TCIMACRO{\U{2102} }%
%BeginExpansion
\mathbb{C}
%EndExpansion
$) in the parameters of the quantum search Hamiltonian $\mathcal{H}$ when the
quantum overlap $x$ approaches zero.

\emph{Case 1}: $\alpha=\delta$, and $\beta=\gamma=0$. In this case, the
Hamiltonian $\mathcal{H}$ in Eq. (\ref{hamilton}) is given by
\begin{equation}
\mathcal{H}_{1}\overset{\text{def}}{=}\alpha E\left[  \left\vert \psi
_{w}\right\rangle \left\langle \psi_{w}\right\vert +\left\vert \psi
_{s}\right\rangle \left\langle \psi_{s}\right\vert \right]  \text{.}
\label{h1}%
\end{equation}
Furthermore, the the maximum value of the transition probability in Eq.
(\ref{nono}) becomes $\mathcal{P}_{\max}=1$ reached at the time
$t_{\mathcal{H}_{1}}^{\ast}$,%
\begin{equation}
t_{\mathcal{H}_{1}}^{\ast}=\frac{1}{\alpha x}\frac{\pi\hslash}{2E}\text{.}
\label{tstar1}%
\end{equation}
Observe that when $\alpha=1$ in\ Eq. (\ref{tstar1}), we recover the well-known
result by Farhi and Guttmann. As a side remark, we point out that
$t_{\mathcal{H}_{1}}^{\ast}$ in Eq. (\ref{tstar1}) is inversely proportional
to the quantum overlap $x$ between the initial state $\left\vert \psi
_{s}\right\rangle $ and the target state $\left\vert \psi_{w}\right\rangle $.

\emph{Case 2}: $\alpha\neq\delta$, and $\beta=\gamma=0$. Using these working
assumptions, the Hamiltonian $\mathcal{H}$ in Eq. (\ref{hamilton}) becomes
\begin{equation}
\mathcal{H}_{2}\overset{\text{def}}{=}E\left[  \alpha\left\vert \psi
_{w}\right\rangle \left\langle \psi_{w}\right\vert +\delta\left\vert \psi
_{s}\right\rangle \left\langle \psi_{s}\right\vert \right]  \text{.}
\label{h2}%
\end{equation}
In this case, $\mathcal{P}_{\max}$ is given by,%
\begin{equation}
\mathcal{P}_{\max}=\frac{\left(  \alpha+\delta\right)  ^{2}x^{2}}{4x^{2}%
\alpha\delta+\left(  \alpha-\delta\right)  ^{2}}\text{.} \label{p2}%
\end{equation}
This maximum $\mathcal{P}_{\max}$ with $0\leq\mathcal{P}_{\max}\leq1$ is
reached at $t_{\mathcal{H}_{2}}^{\ast}$,%
\begin{equation}
t_{\mathcal{H}_{2}}^{\ast}=\frac{2}{\sqrt{4x^{2}\alpha\delta+\left(
\alpha-\delta\right)  ^{2}}}\frac{\pi\hslash}{2E}\text{.} \label{tstar2}%
\end{equation}
Note that $\mathcal{P}_{\max}$ in Eq. (\ref{p2}), when viewed as a function of
$x$, assumes it maximum value $1$ when $\alpha=\delta$. Furthermore,
$\mathcal{P}_{\max}=1$ when $x=1$ for any choice of $\alpha$ and $\delta$.
Furthermore, $t_{\mathcal{H}_{1}}^{\ast}\geq$ $t_{\mathcal{H}_{2}}^{\ast}$
when $0\leq\delta/\left(  1-4x^{2}\right)  \leq\alpha$. In particular, when
$\alpha=\delta/\left(  1-4x^{2}\right)  $, we get%
\begin{equation}
\frac{2E}{\pi\hslash}t_{\mathcal{H}_{1}}^{\ast}=\frac{1-4x^{2}}{\delta
x}=\frac{2E}{\pi\hslash}t_{\mathcal{H}_{2}}^{\ast}\text{.}%
\end{equation}
Finally, we remark that when $0$ $\leq\alpha-\delta\ll1$, the approximate
expression of $\mathcal{P}_{\max}$ in Eq. (\ref{p2}) becomes%
\begin{equation}
\mathcal{P}_{\max}=1-\frac{1}{4}\frac{1-x^{2}}{\alpha^{2}x^{2}}\left(
\alpha-\delta\right)  ^{2}+\mathcal{O}\left(  \left\vert \alpha-\delta
\right\vert ^{3}\right)  \text{.} \label{profound}%
\end{equation}
This approximate maximum transition probability value $\mathcal{P}_{\max}$ in
Eq. (\ref{profound}) is achieved when%
\begin{equation}
t_{\mathcal{H}_{2}}^{\ast}=\left[  \frac{1}{\alpha x}-\frac{1}{8}\frac{\left(
\alpha-\delta\right)  ^{2}}{\alpha^{3}x^{3}}+\mathcal{O}\left(  \left\vert
\alpha-\delta\right\vert ^{3}\right)  \right]  \frac{\pi\hslash}{2E}\text{,}%
\end{equation}
that is, $t_{\mathcal{H}_{2}}^{\ast}=t_{\mathcal{H}_{1}}^{\ast}+\mathcal{O}%
\left(  \left\vert \alpha-\delta\right\vert ^{2}\right)  $.

\begin{figure}[t]
\centering
\includegraphics[width=0.75\textwidth] {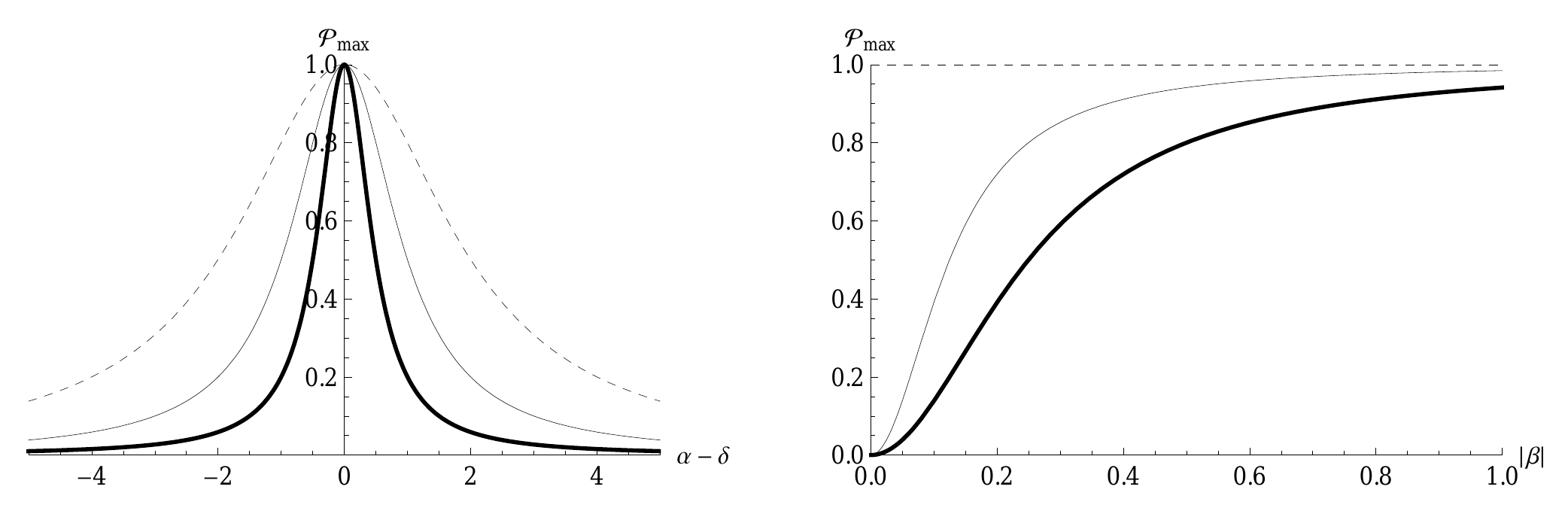}\caption{Maximal success
probability $\mathcal{P}_{\max}$ as a function of $\alpha-\delta$ for
$\left\vert \beta\right\vert =0.25$ (dotted line), $\left\vert \beta
\right\vert =0.5$ (thin solid line), and $\left\vert \beta\right\vert =1$
(thick solid line) in the working assumption that $x$ approaches zero (left);
Maximal success probability $\mathcal{P}_{\max}$ as a function of $\left\vert
\beta\right\vert $ for $\alpha-\delta=0$ (dotted line), $\alpha-\delta=0.25$
(thin solid line), and $\alpha-\delta=0.5$ (thick solid line) in the working
assumption that $x$ approaches zero (right).}%
\label{fig4}%
\end{figure}

\emph{Case 3}: $\beta=\gamma^{\ast}$ \emph{real}, and $\alpha=\delta=0$. In
this case, the Hamiltonian $\mathcal{H}$ in Eq. (\ref{hamilton}) is given by
\begin{equation}
\mathcal{H}_{3}\overset{\text{def}}{=}\beta E\left[  \left\vert \psi
_{w}\right\rangle \left\langle \psi_{s}\right\vert +\left\vert \psi
_{s}\right\rangle \left\langle \psi_{w}\right\vert \right]  \text{.}
\label{h3}%
\end{equation}
The Hamiltonian $\mathcal{H}_{3}$ can be used to search for the target state
$\left\vert \psi_{w}\right\rangle $ with certainty since the maximum
probability value $\mathcal{P}_{\max}$ is given by $\mathcal{P}_{\max}=1$.
This maximum $\mathcal{P}_{\max}$ is reached at $t_{\mathcal{H}_{3}}^{\ast}$,%
\begin{equation}
t_{\mathcal{H}_{3}}^{\ast}\overset{\text{def}}{=}\frac{1}{\beta}\frac
{\pi\hslash}{2E}\text{.} \label{tstar3}%
\end{equation}
Note that, unlike the previous two cases, the time $t_{\mathcal{H}_{3}}^{\ast
}$ does not depend on the quantum overlap $x$.

\emph{Case 4}: $\beta=\gamma^{\ast}$ $\emph{complex}$, and $\alpha=\delta=0$.
Using these working assumptions, the Hamiltonian $\mathcal{H}$ in Eq.
(\ref{hamilton}) becomes
\begin{equation}
\mathcal{H}_{4}\overset{\text{def}}{=}E\left[  \beta\left\vert \psi
_{w}\right\rangle \left\langle \psi_{s}\right\vert +\beta^{\ast}\left\vert
\psi_{s}\right\rangle \left\langle \psi_{w}\right\vert \right]  \text{.}
\label{h4}%
\end{equation}
In this case, $\mathcal{P}_{\max}$ becomes%
\begin{equation}
\mathcal{P}_{\max}=\frac{8\left[  \operatorname{Re}\left(  \beta\right)
\right]  ^{2}x^{2}-4\left[  \operatorname{Re}\left(  \beta\right)  \right]
^{2}x^{4}+4\left\vert \beta\right\vert ^{2}\left(  1-x^{2}\right)  ^{2}%
}{4\left[  \operatorname{Re}\left(  \beta\right)  \right]  ^{2}x^{2}%
+4\left\vert \beta\right\vert ^{2}\left(  1-x^{2}\right)  }\text{.}
\label{pmaxcomplex}%
\end{equation}
This maximum $\mathcal{P}_{\max}$ in\ Eq. (\ref{pmaxcomplex}) is reached at
$t_{\mathcal{H}_{4}}^{\ast}$,%
\begin{equation}
t_{\mathcal{H}_{4}}^{\ast}\overset{\text{def}}{=}\frac{2}{\sqrt{4\left[
\operatorname{Re}\left(  \beta\right)  \right]  ^{2}x^{2}+4\left\vert
\beta\right\vert ^{2}\left(  1-x^{2}\right)  }}\frac{\pi\hslash}{2E}\text{.}%
\end{equation}
Note that, unlike the previous case, the time $t_{\mathcal{H}_{4}}^{\ast}$
does depend on the quantum overlap $x$. Furthermore, observe that if
$\operatorname{Re}\left(  \beta\right)  =0$, $\mathcal{P}_{\max}$ in Eq.
(\ref{pmaxcomplex}) becomes $\mathcal{\tilde{P}}_{\max}=1-x^{2}$. This maximum
$\mathcal{\tilde{P}}_{\max}$ is reached at $\tilde{t}_{\mathcal{H}_{4}}^{\ast
}$,%
\begin{equation}
\tilde{t}_{\mathcal{H}_{4}}^{\ast}\overset{\text{def}}{=}\frac{1}%
{\sqrt{\left\vert \beta\right\vert ^{2}\left(  1-x^{2}\right)  }}\frac
{\pi\hslash}{2E}=\frac{1}{\left\vert \beta\right\vert }\left[  1+\frac{1}%
{2}x^{2}+\mathcal{O}\left(  x^{4}\right)  \right]  \frac{\pi\hslash}%
{2E}=t_{\mathcal{H}_{3}}^{\ast}+\mathcal{O}\left(  x^{2}\right)  \text{.}
\label{ttilda4}%
\end{equation}
In other words, when $0\leq x\ll1$, the search Hamiltonian $\mathcal{H}_{4}$
behaves approximately like the Hamiltonian $\mathcal{H}_{3}$. As a final
remark, we note that when $\beta\overset{\text{def}}{=}2iEx$ we recover
Fenner's quantum search Hamiltonian as proposed in\ Ref. \cite{fenner}.

\emph{Case 5}: $\alpha=\delta$, and $\beta=\gamma^{\ast}$ \emph{real}. In this
case, the Hamiltonian $\mathcal{H}$ in Eq. (\ref{hamilton}) is given by%
\begin{equation}
\mathcal{H}_{5}\overset{\text{def}}{=}\alpha E\left[  \left\vert \psi
_{w}\right\rangle \left\langle \psi_{w}\right\vert +\left\vert \psi
_{s}\right\rangle \left\langle \psi_{s}\right\vert \right]  +\beta E\left[
\left\vert \psi_{w}\right\rangle \left\langle \psi_{s}\right\vert +\left\vert
\psi_{s}\right\rangle \left\langle \psi_{w}\right\vert \right]  \text{.}
\label{h5}%
\end{equation}
It happens that given $\mathcal{H}_{5}$, $\mathcal{P}_{\max}$ becomes
$\mathcal{P}_{\max}=1$. Furthermore, the maximum $\mathcal{P}_{\max}$ is
reached at $t_{\mathcal{H}_{5}}^{\ast}$,%
\begin{equation}
t_{\mathcal{H}_{5}}^{\ast}=\frac{1}{\alpha x+\beta}\frac{\pi\hslash}%
{2E}\text{.} \label{tstar5}%
\end{equation}
Note that for $\beta=0$ and $\alpha=0$, $t_{\mathcal{H}_{5}}^{\ast}$ reduces
to $t_{\mathcal{H}_{1}}^{\ast}$ and $t_{\mathcal{H}_{3}}^{\ast}$,
respectively. For the sake of completeness, we remark that the Hamiltonian
in\ Eq. (\ref{h5}) was originally considered in Ref. \cite{bae02}.

\emph{Case 6}: $\alpha=\delta$, and $\beta=\gamma^{\ast}$ \emph{complex}. In
this case, the Hamiltonian $\mathcal{H}$ in Eq. (\ref{hamilton}) is given by%
\begin{equation}
\mathcal{H}_{6}\overset{\text{def}}{=}\alpha E\left[  \left\vert \psi
_{w}\right\rangle \left\langle \psi_{w}\right\vert +\left\vert \psi
_{s}\right\rangle \left\langle \psi_{s}\right\vert \right]  +E\left[
\beta\left\vert \psi_{w}\right\rangle \left\langle \psi_{s}\right\vert
+\beta^{\ast}\left\vert \psi_{s}\right\rangle \left\langle \psi_{w}\right\vert
\right]  \text{.}%
\end{equation}
Moreover, $\mathcal{P}_{\max}$ becomes%
\begin{equation}
\mathcal{P}_{\max}=\frac{\left\vert \left[  2\operatorname{Re}\left(
\beta\right)  x+2\alpha x^{2}\right]  x+2\left(  \alpha x+\beta\right)
\left(  1-x^{2}\right)  \right\vert ^{2}}{\left[  2\operatorname{Re}\left(
\beta\right)  x+2\alpha x^{2}\right]  ^{2}+4\left(  1-x^{2}\right)  \left[
\left\vert \beta\right\vert ^{2}+2\alpha\operatorname{Re}\left(  \beta\right)
x+\alpha^{2}x^{2}\right]  }\text{.} \label{pmax2}%
\end{equation}
The maximum $\mathcal{P}_{\max}$ is reached at $t_{\mathcal{H}_{6}}^{\ast}$,%
\begin{equation}
t_{\mathcal{H}_{6}}^{\ast}\overset{\text{def}}{=}\frac{2}{\sqrt{\left[
2\operatorname{Re}\left(  \beta\right)  x+2\alpha x^{2}\right]  ^{2}+4\left(
1-x^{2}\right)  \left[  \left\vert \beta\right\vert ^{2}+2\alpha
\operatorname{Re}\left(  \beta\right)  x+\alpha^{2}x^{2}\right]  }}\frac
{\pi\hslash}{2E}\text{.}%
\end{equation}
\begin{table}[t]
\centering
\begin{tabular}
[c]{c|c|c|c|c}%
$\mathcal{H}$ & ${\mathcal{P}}_{\text{max}}$ & $t_{\mathcal{H}}^{\ast}$ &
$(\alpha,\delta)$ & $(\beta,\gamma)$\\\hline
$\mathcal{H}_{1}$ & $1$ & $\frac{\pi\hslash}{2E}(\alpha x)^{-1}$ &
$\alpha=\delta\neq0$ & $\beta=\gamma^{*}=0$\\
$\mathcal{H}_{3}$ & $1$ & $\frac{\pi\hslash}{2E}(\beta)^{-1}$ & $\alpha
=\delta=0$ & $\beta=\gamma^{*}\in\mathbb{R}\backslash\{0\}$\\
$\mathcal{H}_{5}$ & $1$ & $\frac{\pi\hslash}{2E}(\alpha x+\beta)^{-1}$ &
$\alpha=\delta\neq0$ & $\beta=\gamma^{*}\in\mathbb{R}\backslash\{0\}$%
\end{tabular}
\caption{Summary of cases where unit maximal success probability values
$\mathcal{P}_{\max}$ can be achieved for a variety of choices of the
parameters $\alpha$, $\beta$, $\gamma$, and $\delta$ specifying the quantum
search Hamiltonian $\mathcal{H}$ in Eq. (\ref{hamilton}).}%
\end{table}

\emph{Case 7}: $\alpha\neq\delta$, and $\beta=\gamma^{\ast}$ \emph{real}. The
Hamiltonian $\mathcal{H}$ in Eq. (\ref{hamilton}) is given by,
\begin{equation}
\mathcal{H}_{7}\overset{\text{def}}{=}E\left[  \alpha\left\vert \psi
_{w}\right\rangle \left\langle \psi_{w}\right\vert +\delta\left\vert \psi
_{s}\right\rangle \left\langle \psi_{s}\right\vert \right]  +\beta E\left[
\left\vert \psi_{w}\right\rangle \left\langle \psi_{s}\right\vert +\left\vert
\psi_{s}\right\rangle \left\langle \psi_{w}\right\vert \right]  \text{.}%
\end{equation}
In this case, $\mathcal{P}_{\max}$ becomes%
\begin{equation}
\mathcal{P}_{\max}=\frac{\left[  \left(  \alpha+\delta\right)  x+2\beta
\right]  ^{2}}{4\left[  \alpha\delta x^{2}+\left(  \alpha\beta+\beta
\delta\right)  x+\beta^{2}\right]  +\left(  \alpha-\delta\right)  ^{2}%
}\text{.} \label{pm7}%
\end{equation}
The maximum $\mathcal{P}_{\max}$ in Eq. (\ref{pm7}) is reached at
$t_{\mathcal{H}_{7}}^{\ast}$,%
\begin{equation}
t_{\mathcal{H}_{7}}^{\ast}=\frac{2}{\sqrt{4\left[  \alpha\delta x^{2}+\left(
\alpha\beta+\beta\delta\right)  x+\beta^{2}\right]  +\left(  \alpha
-\delta\right)  ^{2}}}\frac{\pi\hslash}{2E}\text{.}%
\end{equation}
Finally, we remark that when $0$ $\leq\alpha-\delta\ll1$, the approximate
expression of $\mathcal{P}_{\max}$ in Eq. (\ref{pm7}) becomes%
\begin{equation}
\mathcal{P}_{\max}=1-\frac{1}{4}\frac{1-x^{2}}{\left(  \alpha x+\beta\right)
^{2}}\left(  \alpha-\delta\right)  ^{2}+\mathcal{O}\left(  \left\vert
\alpha-\delta\right\vert ^{3}\right)  \text{.} \label{app}%
\end{equation}
This approximate maximum transition probability value in Eq. (\ref{app}) is
achieved when%
\begin{equation}
t_{\mathcal{H}_{7}}^{\ast}=\left[  \frac{1}{\alpha x+\beta}-\frac{1}{8}%
\frac{\left(  \alpha-\delta\right)  ^{2}}{\left(  \alpha x+\beta\right)  ^{3}%
}+\mathcal{O}\left(  \left\vert \alpha-\delta\right\vert ^{3}\right)  \right]
\frac{\pi\hslash}{2E}\text{,}%
\end{equation}
that is, $t_{\mathcal{H}_{7}}^{\ast}=t_{\mathcal{H}_{5}}^{\ast}+\mathcal{O}%
\left(  \left\vert \alpha-\delta\right\vert ^{2}\right)  $ with
$t_{\mathcal{H}_{5}}^{\ast}$ in Eq. (\ref{tstar5}).

In Table II we describe the minimum search times $t_{\mathcal{H}_{i}}^{\ast}$
with $i\in\left\{  1\text{, }3\text{, }5\right\}  $ when the maximal success
probability $\mathcal{P}_{\max}$ equals one.\textbf{ }Furthermore,
Fig\textbf{. }$5$ displays two plots. The plot on the left represents the
minimum search time $t^{\ast}$ versus the overlap $x$ assuming $\alpha
=\beta=1$ and $E=h=1$. From this plot, we note that $t_{\mathcal{H}_{5}}%
^{\ast}\leq t_{\mathcal{H}_{3}}^{\ast}\leq t_{\mathcal{H}_{1}}^{\ast}$. The
plot on the right, instead, represents the temporal behavior of the success
probability $\mathcal{P}\left(  t\right)  $ assuming $\alpha=\beta=1$,
$E=h=1$,\textbf{ }and $x=0.5$. We observe that $\mathcal{P}\left(  t\right)  $
reaches the ideal unit value with $\mathcal{H}_{5}$ at $t_{\mathcal{H}_{5}%
}^{\ast}=1/6\simeq0.17$, with $\mathcal{H}_{3}$ at $t_{\mathcal{H}_{3}}^{\ast
}=1/4=0.25$\textbf{, }and with $\mathcal{H}_{1}$ at $t_{\mathcal{H}_{1}}%
^{\ast}=1/2=0.5$. Despite the detrimental effects of asymmetries and
complexities on the achievable maximal success probability values represented
in Fig\textbf{.} $4$ when $x$ approaches zero and despite the fact as reported
in Table II and Fig. $5$ that $\mathcal{H}_{5}$ appears to be the quantum
search Hamiltonian that yields the shortest search time needed to achieve unit
success probability, we point out that it is possible to suitably choose the
Hamiltonian parameters $\alpha$, $\beta$, $\gamma$, and $\delta$ in
$\mathcal{H}$ together with the overlap $x$ in such a manner that nearly
optimal success probability threshold values can be obtained in search times
shorter than those specified by $\mathcal{H}_{5}$. Indeed, Fig. $6$ displays
such a circumstance. Assuming $\alpha=\delta=0.5$, $\beta=1$, and $x=0.5$, the
unit success probability is obtained with $\mathcal{H}_{5}$ at $t_{\mathcal{H}%
_{5}}^{\ast}=1/5=0.2$ while the chosen threshold value $\mathcal{P}%
_{\text{threshold}}=0.95$ is reached at $\tilde{t}_{\mathcal{H}_{5}}%
\simeq0.1667$. However, assuming $\mathcal{H}$ with $\alpha=0.5$, $\delta=1$,
$\beta=1$, and $x=0.5$, despite the fact that the maximal success probability
is only nearly optimal with $\mathcal{P}_{\max}\simeq0.9758\leq1$,\textbf{
}the selected threshold value $\mathcal{P}_{\text{threshold}}=0.95$\textbf{
}is reached at $\tilde{t}_{\mathcal{H}}\simeq0.1579\leq\tilde{t}%
_{\mathcal{H}_{5}}$. For a discussion on the choice of the numerical values of
the quantum overlap $x$, we refer to Appendix C.

\begin{figure}[t]
\centering
\includegraphics[width=0.75\textwidth] {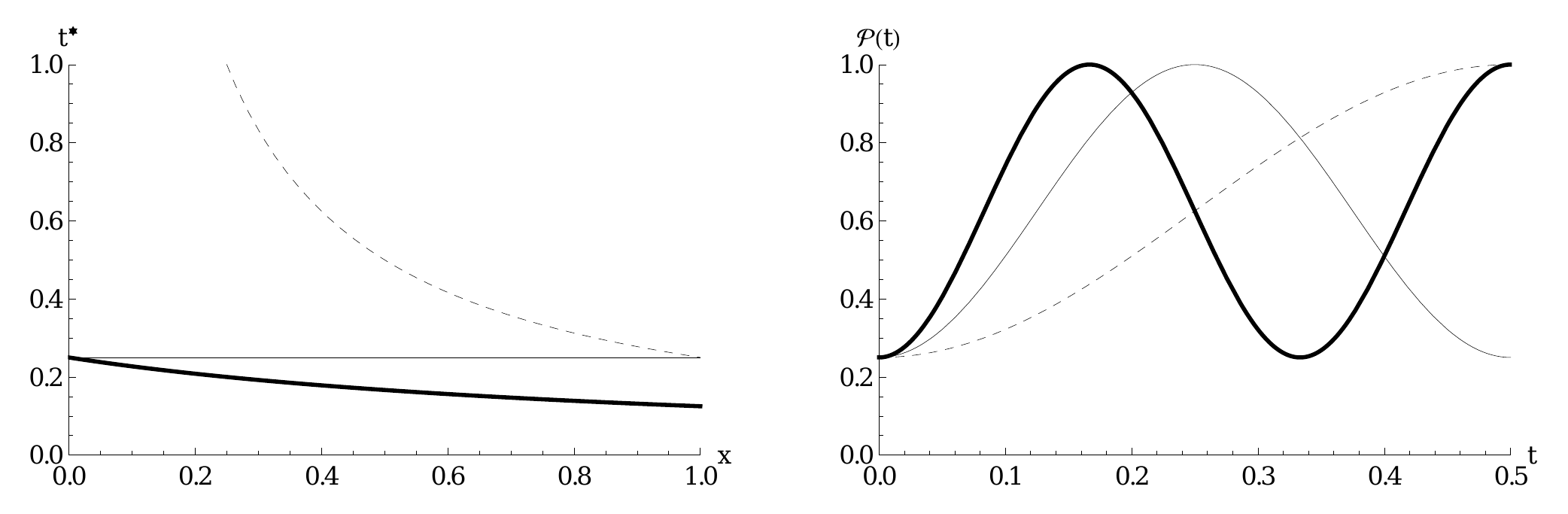}\caption{The plot on the left
displays the minimum search time $t^{\ast}$ versus the quantum overlap $x$ for
the search Hamiltonian $\mathcal{H}_{1}$ (dashed line), $\mathcal{H}_{3}$
(thin solid line), and $\mathcal{H}_{5}$ (thick solid line). The plot on the
right, instead, shows $\mathcal{P}\left(  t\right)  $ versus $t$ for the
search Hamiltonian $\mathcal{H}_{1}$ (dashed line), $\mathcal{H}_{3}$ (thin
solid line), and $\mathcal{H}_{5}$ (thick solid line). In the former plot, we
assume $\alpha=\beta=1$ and $E=h=1$. In the latter plot, we also assume
$x=1/2$.}%
\label{fig5}%
\end{figure}

\section{Concluding Remarks}

In this paper, we presented a detailed analysis concerning the computational
aspects needed to analytically evaluate the transition probability from a
source state $\left\vert s\right\rangle $ to a target state $\left\vert
w\right\rangle $ in a continuous time quantum search problem defined by a
multi-parameter generalized time-independent Hamiltonian $\mathcal{H}$ in\ Eq.
(\ref{hamilton}). In particular, quantifying the performance of a quantum
search in terms of speed (minimum search time, $t^{\ast}$) and fidelity (high
success probability, $\mathcal{P}$), we consider a variety of special cases
that emerge from the generalized Hamiltonian. Finally, recovering also the
well-known Farhi-Gutmann analog quantum search scenario, we briefly discuss
the relevance of a tradeoff between speed and fidelity with emphasis on issues
of both theoretical and practical importance to quantum information processing.

\subsection{Summary of main results}

Our main conclusions can be summarized as follows.

\begin{enumerate}
\item[{[1]}] First, we provided a detailed analytical computation of the
transition probability $\mathcal{P}_{\left\vert s\right\rangle \rightarrow
\left\vert w\right\rangle }\left(  t\right)  $ in Eq. (\ref{it}) from the
source state $\left\vert s\right\rangle $ to the target state $\left\vert
w\right\rangle $ under the working assumption that the quantum mechanical
evolution is governed by the generalized quantum search Hamiltonian
$\mathcal{H}$. Such a computation, despite being straightforward, is quite
tedious. Therefore, we have reason to believe it can be relevant to the novice
with a growing interest in analog quantum search algorithms as well as to the
expert seeking to find nearly-optimal solutions in realistic quantum search
problems where a tradeoff between fidelity and minimum search time is required;

\item[{[2]}] Second, given the family $\mathcal{F}_{\mathcal{H}}%
\overset{\text{def}}{=}\left\{  \mathcal{H}\right\}  $ with $\mathcal{H}%
=\mathcal{H}\left(  x\text{; }\alpha\text{, }\beta\text{, }\gamma\text{,
}\delta\right)  $ where $\alpha$ and $\beta\in%
%TCIMACRO{\U{211d} }%
%BeginExpansion
\mathbb{R}
%EndExpansion
$ while $\beta=\gamma^{\ast}$ $\in%
%TCIMACRO{\U{2102} }%
%BeginExpansion
\mathbb{C}
%EndExpansion
$, we have conveniently identified two sub-families $\mathcal{F}_{\mathcal{H}%
}^{\left(  \text{optimal}\right)  }\overset{\text{def}}{=}\left\{
\mathcal{H}_{1}\text{, }\mathcal{H}_{3}\text{, }\mathcal{H}_{5}\right\}  $ and
$\mathcal{F}_{\mathcal{H}}^{\left(  \text{nearly-optimal}\right)  }%
\overset{\text{def}}{=}\left\{  \mathcal{H}_{2}\text{, }\mathcal{H}_{4}\text{,
}\mathcal{H}_{6}\text{, }\mathcal{H}_{7}\right\}  $ that contain quantum
search Hamiltonians yielding optimal and nearly-optimal fidelity values,
respectively. The former sub-family is specified by the asymmetry between the
\emph{real} parameters $\alpha$ and $\delta$. The latter sub-family, instead,
is characterized by the complexity (that is, the essence of being
\emph{complex}-valued) of the parameters $\beta$ and $\gamma$. Each element of
the family has been classified with respect to its maximal success probability
and the minimum time necessary to reach such a value.\ An overview of these
results appears in Table I. In addition, in Fig. $4$ we report on the
detrimental effects caused by the presence of asymmetries and complexities in
the parameters that specify the particular quantum search Hamiltonian on the
maximal success probability in the limiting working assumption that the source
state and the target state are orthogonal.

\item[{[3]}] Third, we ranked the performance of each element of the
sub-family $\mathcal{F}_{\mathcal{H}}^{\left(  \text{optimal}\right)  }$ by
analyzing the minimum search time required to reach unit fidelity. These
results are displayed in Table II. In particular, as evident from Fig. $5$, we
find\textbf{ }that $\mathcal{H}_{5}$ can outperform the Farhi-Gutmann search
Hamiltonian $\mathcal{H}_{1}$ in terms of speed.

\item[{[4]}] Lastly, despite the observed detrimental effects of asymmetries
and complexities on the numerical values of the maximal success probabilities,
we find that imperfect search Hamiltonians can outperform perfect search
Hamiltonians provided that only a large nearly-optimal fidelity value is
sought. This finding is reported in Fig. $6$.
\end{enumerate}

\subsection{Limitations and possible future developments}

In what follows, we report some limitations together with possible future
improvements of our investigation.

\begin{enumerate}
\item[{[1]}] First, we have reason to believe our analysis presented in this
paper could be a useful starting point for a more rigorous investigation that
would include both experimental and theoretical aspects of a tradeoff between
fidelity and run time in quantum search algorithms. Indeed, we are aware that
it is helpful to decrease the control time of the control fields employed to
generate a target quantum state or a target quantum gate in order to mitigate
the effect of decoherence originating from the interaction of a quantum system
with the environment. Moreover, we also know that it may be convenient to
increase the control time beyond a certain critical value to enhance the
fidelity of generating such targets and reach values arbitrarily close to the
maximum $\mathcal{F}=1$. However, when the control time reaches a certain
value that may be close to the critical value, decoherence can become a
dominant effect. Therefore, investigating the tradeoff between time control
and fidelity can be of great practical importance in quantum computing
\cite{rabitz12,rabitz15,cappellaro18}. Given that it is very challenging to
find a rigorous optimal time control and in many cases the control is only
required to be sufficiently precise and short, one can design algorithms
seeking suboptimal control solutions for much reduced computational effort.
For instance, the fidelity of tomography experiments is rarely above $99\%$
due to the limited control precision of the tomographic experimental
techniques as pointed out in Ref. \cite{rabitz15}. Under such conditions, it
is unnecessary to prolong the control time since the departure from the
optimal scenario is essentially negligible. Hence, it can certainly prove
worthwhile to design slightly suboptimal algorithms that can be much cheaper computationally.

\item[{[2]}] Second, we speculate it may be worth pursuing the possibility of
borrowing ideas from approximate quantum cloning to design approximate quantum
search algorithms capable of finding targets in the presence of\textbf{
}\emph{a priori }information. As a matter of fact, recall that the no-cloning
theorem in quantum mechanics states that it is impossible to consider a
cloning machine capable of preparing two exact copies of a completely unknown
pure qubit state \cite{zurek82}. However, with the so-called universal cloner
\cite{hillery96} (that is, a state-independent symmetric cloner) acting on the
whole Bloch sphere, it is possible to prepare two approximate copies of an
unknown pure qubit state with the same fidelity\textbf{ }$\mathcal{F}%
=5/6<1$\textbf{. }Interestingly, it is possible to enhance these fidelity
values achieved with a universal cloner by properly exploiting some
relevant\textbf{ }\emph{a priori}\textbf{ }information on a given quantum
state that one wishes to clone. This idea of exploiting\textbf{ }\emph{a
priori}\textbf{ }information generated a number of state-dependent cloning
machines capable of outperforming the universal cloner for some special set of
qubits. For instance, phase-covariant cloners are especially successful for
cloning qubits chosen from the equator of the Bloch sphere
\cite{macchiavello00} while belt quantum cloning machines are very efficient
in cloning qubits between two latitudes on the Bloch sphere \cite{wang09}. For
an interesting method for improving the cloning fidelity in terms of\textbf{
}\emph{a priori}\textbf{ }amplitude and phase information, we refer to Ref.
\cite{kang16}. We shall investigate this line of investigation in forthcoming efforts.

\item[{[3]}] Third, from a more applied perspective, despite its relative
simplicity, the idea of finding a tradeoff between search time and fidelity in
analog quantum searching as presented in this paper could be potentially
regarded as a valid starting point for a time-fidelity tradeoff analysis in
disease diagnosis in complex biological systems. For these systems, the source
and target states are replaced with the source and target patterns,
respectively. In particular, the target pattern classifies the type of illness
being searched. For recent important investigations based upon the joint use
of quantum field theoretic methods and general relativity techniques
concerning the transition from source to target patterns in complex biological
systems, including DNA\ and protein structures, we refer to Refs.
\cite{capozziello1,capozziello2}. More realistic applications of our work are
very important and we shall also give a closer look to these aspects in the
near future.

\item[{[4]}] Fourth, a further possibility could be related to cosmology. As
discussed in
\cite{capozziello3,capozziello4,luongo19,capozziello13,capozziello11}, there
exist possible connections between quantum entanglement and cosmological
observational parameters. In fact, assuming that two cosmological epochs are
each other entangled, by measuring the entanglement degree, it is possible to
recover dynamical properties. Specifically, the effects of the so called
\textit{dark energy} could be due to the entanglement between states, since a
negative pressure arises. In this process, an \textquotedblleft entanglement
weight\textquotedblright, the so-called negativity of entanglement can be
defined and then the apparent today observed accelerated expansion occurs when
the cosmological parameters are entangled. In this perspective, dark energy
could be seen as a straightforward consequence of entanglement without
invoking (not yet observed) further material fundamental components. The
present analysis could help in this cosmological perspective once the
cosmological equations are modeled out as Schr\"{o}dinger-like equations as
discussed in \cite{capozziello5}.

\item[{[5]}] Lastly, in real life scenarios, searching in a completely
unstructured fashion can be unnecessary. Instead, the search can be guided by
employing some \emph{prior} relevant information about the location of the
target state. Interestingly, this is what happens in the framework of quantum
search with advice \cite{montanaro11,montanaro17}. In this framework, the aim
is to minimize the expected number of queries with respect to the probability
distribution encoding relevant information about where the target might be
located. A major advancement in the work we presented in this paper would be
figuring out a systematic way to incorporate relevant \emph{prior} information
about the possible location of the target directly into the continuous time
quantum search Hamiltonian. We leave this intriguing open problem to future
scientific endeavours.
\end{enumerate}

In conclusion, our proposed analysis was partially inspired by some of our
previous findings reported in Refs. \cite{cafaro-alsing19A, cafaro-alsing19B}
and can be improved in a number of ways in the immediate short term. For
instance, it would be interesting to address the following question: How large
should the nearly optimal fidelity value be chosen, given the finite precision
of quantum measurements and the unavoidable presence of decoherence effects in
physical implementations of quantum information processing tasks? We leave the
investigation of a realistic tradeoff between speed and fidelity in analog
quantum search problems to forthcoming scientific efforts.

\begin{figure}[t]
\centering
\includegraphics[width=0.35\textwidth] {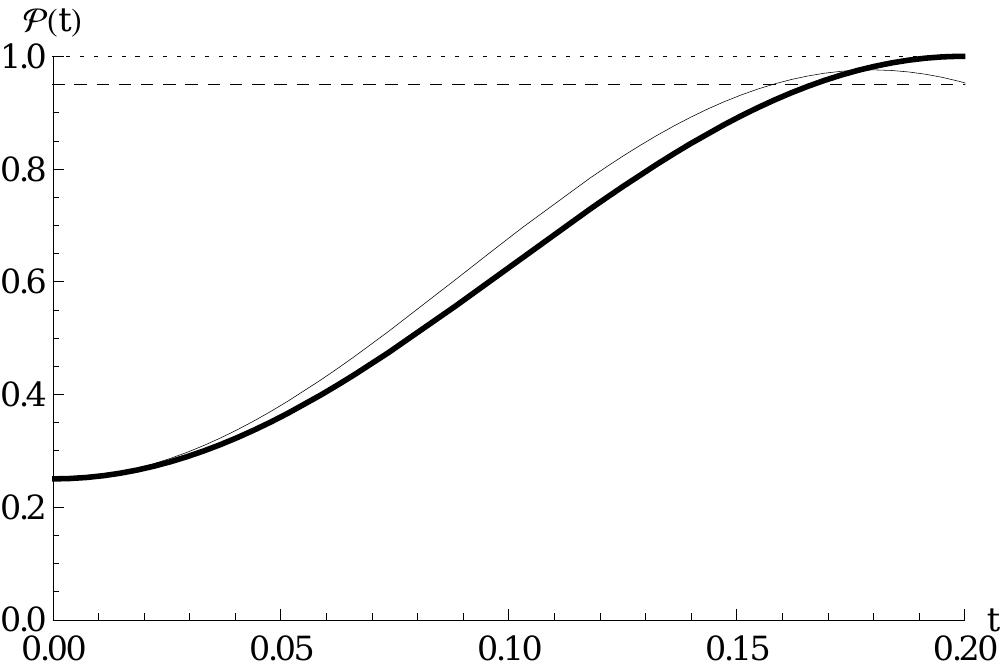}\caption{The thin and the thick
solid lines display $\mathcal{P}\left(  t\right)  $ versus $t$ for the search
Hamiltonians $\mathcal{H}$ and $\mathcal{H}_{5}$, respectively. In the former
case, we set $\alpha=0.5$, $\delta=1$, $\beta=1$, and $x=0.5$. In the latter
case, instead, we set $\alpha=\delta=0.5$, $\beta=1$, and $x=0.5$. In both
cases, we also assume $E=h=1$. The dashed line denotes the chosen threshold
success probability value $\mathcal{P}_{\text{threshold}}=0.95$. Finally, the
dotted line denotes the optimal success probability $\mathcal{P}=1$.}%
\label{fig6}%
\end{figure}

\begin{acknowledgments}
C. C. acknowledges the hospitality of the United States Air Force Research
Laboratory (AFRL) in Rome-NY where part of his contribution to this work was
completed. S.C. acknowledges partial support of \textit{Istituto Italiano di
Fisica Nucleare} (INFN), \textit{iniziative specifiche} QGSKY and MOONLIGHT2.
\end{acknowledgments}

\bigskip\pagebreak

\appendix

\section{Reality of eigenvalues of the Hamiltonian in\ Eq. (\ref{hamilton})}

In this Appendix, we briefly show the well-known fact that if an operator $T$
is Hermitian, then its eigenvalues $\left\{  \lambda_{n}\right\}  $ are\emph{
real} and its eigenvectors $\left\{  \left\vert v_{\lambda_{n}}\right\rangle
\right\}  $ corresponding to distinct eigenvalues are orthogonal. Indeed,
assuming that $\left\{  \left\vert v_{\lambda_{n}}\right\rangle \right\}  $
are normalized and $T\left\vert v_{\lambda_{n}}\right\rangle =\lambda
_{n}\left\vert v_{\lambda_{n}}\right\rangle $ with $T=T^{\dagger}$, we have%
\begin{align}
\lambda_{n}  &  =\frac{\langle v_{\lambda_{n}}|T|v_{\lambda_{n}}\rangle
}{\langle v_{\lambda_{n}}|v_{\lambda_{n}}\rangle}=\frac{\left(  \langle
v_{\lambda_{n}}|T^{\dagger}|v_{\lambda_{n}}\rangle\right)  ^{\ast}}{\langle
v_{\lambda_{n}}|v_{\lambda_{n}}\rangle}\nonumber\\
&  =\frac{\left(  \langle v_{\lambda_{n}}|T|v_{\lambda_{n}}\rangle\right)
^{\ast}}{\langle v_{\lambda_{n}}|v_{\lambda_{n}}\rangle}=\frac{\left(
\lambda_{n}\langle v_{\lambda_{n}}|v_{\lambda_{n}}\rangle\right)  ^{\ast}%
}{\langle v_{\lambda_{n}}|v_{\lambda_{n}}\rangle}\nonumber\\
&  =\frac{\lambda_{n}^{\ast}\langle v_{\lambda_{n}}|v_{\lambda_{n}}\rangle
}{\langle v_{\lambda_{n}}|v_{\lambda_{n}}\rangle}=\lambda_{n}^{\ast}\text{,}%
\end{align}
that is, $\lambda_{n}=\lambda_{n}^{\ast}\in%
%TCIMACRO{\U{211d} }%
%BeginExpansion
\mathbb{R}
%EndExpansion
$. Furthermore, given $i\neq j$ with $1\leq i$, $j\leq n$, assume $T\left\vert
v_{\lambda_{i}}\right\rangle =\lambda_{i}\left\vert v_{\lambda_{i}%
}\right\rangle $ and $T\left\vert v_{\lambda_{j}}\right\rangle =\lambda
_{j}\left\vert v_{\lambda_{j}}\right\rangle $. Then, we obtain%
\begin{equation}
\lambda_{j}\langle v_{\lambda_{i}}|v_{\lambda_{j}}\rangle=\langle
v_{\lambda_{i}}|T|v_{\lambda_{j}}\rangle=\langle v_{\lambda_{i}}|T^{\dagger
}|v_{\lambda_{j}}\rangle=\lambda_{i}^{\ast}\langle v_{\lambda_{i}}%
|v_{\lambda_{j}}\rangle=\lambda_{i}\langle v_{\lambda_{i}}|v_{\lambda_{j}%
}\rangle\text{,}%
\end{equation}
that is,%
\begin{equation}
\left(  \lambda_{i}-\lambda_{j}\right)  \langle v_{\lambda_{i}}|v_{\lambda
_{j}}\rangle=0\text{.} \label{orthogonal}%
\end{equation}
Since $\lambda_{i}\neq\lambda_{j}$ by assumption, Eq. (\ref{orthogonal})
implies that $\left\vert v_{\lambda_{i}}\right\rangle $ and $|v_{\lambda_{j}%
}\rangle$ must be orthogonal.

\section{Derivation of Eq. (\ref{fess})}

In this Appendix we derive Eq. (\ref{fess}). Specifically, exploiting standard
trigonometric relations in a clever sequential order, we obtain%
\begin{align}
\mathcal{P}\left(  \tilde{\alpha}\right)   &  =\left\vert \tilde{A}\right\vert
^{2}+\left\vert \tilde{B}\right\vert ^{2}+2\tilde{A}\tilde{B}^{\ast}%
\cos\left(  2\tilde{\alpha}\right) \nonumber\\
&  =\left\vert \tilde{A}\right\vert ^{2}+\left\vert \tilde{B}\right\vert
^{2}+2\tilde{A}\tilde{B}^{\ast}\left[  \cos^{2}\left(  \tilde{\alpha}\right)
-\sin^{2}\left(  \tilde{\alpha}\right)  \right] \nonumber\\
&  =\left\vert \tilde{A}\right\vert ^{2}+\left\vert \tilde{B}\right\vert
^{2}+2\tilde{A}\tilde{B}^{\ast}\cos^{2}\left(  \tilde{\alpha}\right)
-2\tilde{A}\tilde{B}^{\ast}\sin^{2}\left(  \tilde{\alpha}\right) \nonumber\\
&  =\left\vert \tilde{A}\right\vert ^{2}\sin^{2}\left(  \tilde{\alpha}\right)
+\left\vert \tilde{A}\right\vert ^{2}\cos^{2}\left(  \tilde{\alpha}\right)
+\left\vert \tilde{B}\right\vert ^{2}\sin^{2}\left(  \tilde{\alpha}\right)
+\left\vert \tilde{B}\right\vert ^{2}\cos^{2}\left(  \tilde{\alpha}\right)
+2\tilde{A}\tilde{B}^{\ast}\cos^{2}\left(  \tilde{\alpha}\right)  -2\tilde
{A}\tilde{B}^{\ast}\sin^{2}\left(  \tilde{\alpha}\right) \nonumber\\
&  =\left(  \left\vert \tilde{A}\right\vert ^{2}+\left\vert \tilde
{B}\right\vert ^{2}-2\tilde{A}\tilde{B}^{\ast}\right)  \sin^{2}\left(
\tilde{\alpha}\right)  +\left(  \tilde{A}^{2}+\tilde{B}^{2}+2\tilde{A}%
\tilde{B}^{\ast}\right)  \cos^{2}\left(  \tilde{\alpha}\right) \nonumber\\
&  =\left\vert \tilde{A}-\tilde{B}\right\vert ^{2}\sin^{2}\left(
\tilde{\alpha}\right)  +\left\vert \tilde{A}+\tilde{B}\right\vert ^{2}\cos
^{2}\left(  \tilde{\alpha}\right)  \text{,}%
\end{align}
that is,%
\begin{equation}
\mathcal{P}\left(  \tilde{\alpha}\right)  =\left\vert \tilde{A}-\tilde
{B}\right\vert ^{2}\sin^{2}\left(  \tilde{\alpha}\right)  +\left\vert
\tilde{A}+\tilde{B}\right\vert ^{2}\cos^{2}\left(  \tilde{\alpha}\right)
\text{.}%
\end{equation}
\begin{table}[t]
\centering
\begin{tabular}
[c]{c|c|c|c}%
$N$ & $\sigma_{\theta}^{2}$ & $\text{\textrm{Prob}}^{(\text{uniform})}%
(x\geq\bar{x})$ & $\text{\textrm{Prob}}^{(\text{non-uniform})}(x\geq\bar{x}%
)$\\\hline
$4$ & $10^{-1}$ & $3.72 \times10^{-4}$ & $6.86 \times10^{-3}$\\
$4$ & $1$ & $3.72 \times10^{-4}$ & $14.55 \times10^{-2}$\\
$4$ & $10$ & $3.72 \times10^{-4}$ & $19.07 \times10^{-2}$\\
$8$ & $10^{-1}$ & $1.20 \times10^{-7}$ & $6.91 \times10^{-3}$\\
$8$ & $1$ & $1.20 \times10^{-7}$ & $14.72 \times10^{-2}$\\
$8$ & $10$ & $1.20 \times10^{-7}$ & $19.28 \times10^{-2}$\\
$16$ & $10^{-1}$ & $1.79 \times10^{-14}$ & $6.92 \times10^{-3}$\\
$16$ & $1$ & $1.79 \times10^{-14}$ & $14.74 \times10^{-2}$\\
$16$ & $10$ & $1.79 \times10^{-14}$ & $19.31 \times10^{-2}$\\
&  &  &
\end{tabular}
\caption{Numerical estimates of \textrm{Prob}$\left(  x\geq\bar{x}\right)  $
with $\bar{x}=\cos\left(  \pi/8\right)  \simeq0.92$ assuming the quantum
overlap $x$ both uniformly and non-uniformly distributed. In the latter case,
we assume $\mu_{\theta}=(3/8)\pi$ in the expression of the probability density
function $\rho_{w}\left(  \theta\text{; }N\text{, }\mu_{\theta}\text{, }%
\sigma_{\theta}^{2}\right)  $ in Eq. (\ref{ass}). }%
\end{table}

\section{Numerical values of the quantum overlap}

In this Appendix, we discuss the choice of numerical values of the quantum
overlap $x$ by considering non-uniform probability densities of target states
on the Bloch sphere \cite{nielsen}. The discussion follows closely the
presentation that appears in Ref. \cite{cafaro-alsing19A}.

The Bloch sphere representation of a normalized $n$-qubit state in the Hilbert
space $\mathcal{H}_{2}^{n}$ with $\dim_{%
%TCIMACRO{\U{2102} }%
%BeginExpansion
\mathbb{C}
%EndExpansion
}\left[  \mathcal{H}_{2}^{n}\right]  =N$ with $N\overset{\text{def}}{=}2^{n}$
and $\dim_{%
%TCIMACRO{\U{211d} }%
%BeginExpansion
\mathbb{R}
%EndExpansion
}\left[  \mathcal{H}_{2}^{n}\right]  =2\cdot2^{n}=2^{n+1}=2N$ is given in
terms of a $\left(  2N-1\right)  $-dimensional unit sphere. For a single-qubit
quantum state, for instance, $N=2$ and the Bloch sphere is a $3$-dimensional
unit sphere.

Recall that the space enclosed by a $\left(  2N-1\right)  $-dimensional unit
sphere $\mathcal{S}^{2N-1}$ is a $2N$-ball whose infinitesimal volume element
can be written in spherical coordinates as,%
\begin{equation}
dV_{2N\text{-ball}}^{\left(  \text{spherical}\right)  }\overset{\text{def}}%
{=}r^{2N-1}\sin^{2N-2}\left(  \theta_{1}\right)  \sin^{2N-3}\left(  \theta
_{2}\right)  ...\sin\left(  \theta_{2N-2}\right)  drd\theta_{1}d\theta
_{2}...d\theta_{2N-2}d\theta_{2N-1}\text{,} \label{nball}%
\end{equation}
where $\theta_{i}\in\left[  0\text{, }\pi\right)  $ for any $1\leq i\leq2N-2$
and $\theta_{2N-1}\in\left[  0\text{, }2\pi\right)  $. Furthermore, the volume
element $dV_{\mathcal{S}^{2N-1}}^{\left(  \text{spherical}\right)  }$ of the
$\left(  2N-1\right)  $-dimensional unit sphere generalizes the concept of the
area element of a two-dimensional unit sphere and, from Eq. (\ref{nball}), is
given by%
\begin{equation}
dV_{\mathcal{S}^{2N-1}}^{\left(  \text{spherical}\right)  }\overset
{\text{def}}{=}\sin^{2N-2}\left(  \theta_{1}\right)  \sin^{2N-3}\left(
\theta_{2}\right)  ...\sin\left(  \theta_{2N-2}\right)  drd\theta_{1}%
d\theta_{2}...d\theta_{2N-2}d\theta_{2N-1}\text{.} \label{area}%
\end{equation}
The assumption that the target state $\left\vert w\right\rangle $ is selected
at random on the Bloch sphere means that we assume absolute ignorance (that
is, maximum entropy) about the location of the target. In such a case, the
probability density function (pdf) $\rho_{w}^{\left(  \text{uniform}\right)
}\left(  \theta_{1}\text{,..., }\theta_{2N-1}\right)  $ of the target state
$\left\vert w\right\rangle $ can be chosen to be uniform on the Bloch sphere.
However, if one becomes aware of an important piece of information about the
location of the target, one can think of updating his/her state of knowledge
(see Ref. \cite{cafaropre}, for instance) about the target with a new
(non-uniform) pdf $\rho_{w}^{\left(  \text{non-uniform}\right)  }\left(
\theta_{1}\text{,..., }\theta_{2N-1}\right)  $ of the target state on the
$\left(  2N-1\right)  $-dimensional unit sphere $\mathcal{S}^{2N-1}$. As a
consequence, the search algorithm can be adapted to the target in order to
improve the efficiency of the searching scheme. We emphasize that these
considerations are reminiscent of what happens in channel-adapted quantum
error correction \cite{cafaro14,fletcher07,cafaroosid} and adaptive quantum
computing \cite{briegel15} where classical learning techniques can be used to
enhance the performance of certain quantum tasks \cite{briegel16,briegel18}.
For the time-being, returning to our discussion, we assume that the
probability density function $\rho_{w}^{\left(  \text{non-uniform}\right)
}\left(  \theta_{1}\text{,..., }\theta_{2N-1}\right)  $ will be denoted as
$\rho_{w}\left(  \theta\right)  $ and will depend only the coordinate
$\theta_{1}\overset{\text{def}}{=}\theta$, while it is uniform with respect to
the remaining coordinates $\theta_{2}$,..., $\theta_{2N-1}$. \ Therefore,
by\textbf{\ }marginalizing over all the unimportant integration variables but
$\theta$ with $x\overset{\text{def}}{=}\left\vert \left\langle
s|w\right\rangle \right\vert =\cos\left(  \theta\right)  $ where $0\leq
\theta\leq\pi/2$, we find that the probability that $x$ is greater than a
given value $\bar{x}$ is given by,
\begin{equation}
\text{\textrm{Prob}}\left(  x\geq\bar{x}\right)  \overset{\text{def}}{=}%
\frac{\int_{0}^{\cos^{-1}\left(  \bar{x}\right)  }\rho_{w}\left(
\theta\right)  \left[  \sin\left(  \theta\right)  \right]  ^{2N-2}d\theta
}{\int_{0}^{\frac{\pi}{2}}\rho_{w}\left(  \theta\right)  \left[  \sin\left(
\theta\right)  \right]  ^{2N-2}d\theta}\text{.} \label{prob}%
\end{equation}
The quantity $\rho_{w}\left(  \theta\right)  $ in Eq. (\ref{prob}) denotes a
well-defined pdf, that is to say, a pdf that is positive and normalized to
one. For the sake of reasoning, we select $\left\vert s\right\rangle $ to be
at the north pole of the the $\left(  2N-1\right)  $-dimensional unit sphere
$\mathcal{S}^{2N-1}$. Generalizations to less peculiar scenarios may be
considered in a relatively straightforward manner.

In what follows, we provide some rationale for our selection of $\rho
_{w}\left(  \theta\right)  $. The functional form of $\rho_{w}\left(
\theta\right)  $ is essentially that of a Gaussian with mean $\mu_{\theta}$
and variance $\sigma_{\theta}^{2}$ multiplied by a suitably chosen oscillatory
function. The mean is set equal zero, while any value of $\mu_{\theta}$
between $0$ and $\pi/2$ can be chosen provided that the variance
$\sigma_{\theta}^{2}$ is not too small. Furthermore, the multiplicative factor
in the proposed expression of $\rho_{w}\left(  \theta\right)  $ is chosen in
such a manner as\textbf{\ }to substantially mitigate the oscillatory behavior
of $\left[  \sin\left(  \theta\right)  \right]  ^{2N-2}$ in Eq. (\ref{prob})
and leads when multiplied with it, to an approximately constant function over
the selected domain of integration. Practically, one can consider a narrowly
distributed Gaussian peaked nearby the location of the initial state or, for a
Gaussian peaked far away from such a location, the width of the Gaussian has
to be selected suitably larger. Under this working assumption, a convenient
choice for our analysis is given by the following pdf $\rho_{w}\left(
\theta\right)  =\rho_{w}\left(  \theta\text{; }N\text{, }\mu_{\theta}\text{,
}\sigma_{\theta}^{2}\right)  $,%
\begin{equation}
\rho_{w}\left(  \theta\text{; }N\text{, }\mu_{\theta}\text{, }\sigma_{\theta
}^{2}\right)  \overset{\text{def}}{=}\frac{\mathcal{N}\left(  N\text{, }%
\mu_{\theta}\text{, }\sigma_{\theta}^{2}\right)  \cdot\exp\left(
-\frac{\left(  \theta-\mu_{\theta}\right)  ^{2}}{2\sigma_{\theta}^{2}}\right)
}{1+\left[  10\sin\left(  \theta\right)  \right]  ^{2N-2}}\text{.} \label{ass}%
\end{equation}
In Eq. (\ref{ass}), $\mathcal{N}=\mathcal{N}\left(  N\text{, }\mu_{\theta
}\text{, }\sigma_{\theta}^{2}\right)  $ is a normalization factor that depends
upon the choice of $N$, $\mu_{\theta}$, and $\sigma_{\theta}^{2}$. Note that
the $N$-dependence of $\rho_{w}\left(  \theta\text{; }N\text{, }\mu_{\theta
}\text{, }\sigma_{\theta}^{2}\right)  $ is not uncommon. For instance, a
uniform pdf on an $N$-dimensional hypercube with a side of length $L$ scales
as $L^{-N}$. For the sake of reasoning, assume $N=16$, $\mu_{\theta}=3\pi/8$,
$\sigma_{\theta}^{2}=1$, and $\bar{\theta}=\cos^{-1}\left(  \bar{x}\right)
=\pi/8$ with $\bar{x}=\cos\left(  \pi/8\right)  \simeq0.92$ (large overlap).
In this case, we numerically find that while \textrm{Prob}$^{\left(
\text{uniform}\right)  }\left(  x\geq\bar{x}\right)  \simeq1.79\times10^{-14}$
is essentially zero, \textrm{Prob}$^{\left(  \text{non-uniform}\right)
}\left(  x\geq\bar{x}\right)  \simeq0.15$ becomes non-negligible.\textbf{ }In
Table III we report numerical estimates of the probabilities of $x$ being
greater than a given value $\bar{x}$ for a variety of dimensions\textbf{ }%
$N$\textbf{ }of the Hilbert space and for different features of the
non-uniform probability density function\textbf{ }$\rho_{w}$\textbf{ }of the
target state. We note that the probability of occurrence of high values of $x$
can become non-negligible for suitably chosen non-uniformities in the pdf.
Furthermore, non-uniformities (that is, knowledge of relevant information
about the target state) appear to become more relevant in higher-dimensional
Hilbert spaces.

\end{document}